\documentclass[sigconf,nonacm]{acmart}

\renewcommand\footnotetextcopyrightpermission[1]{} 
\AtBeginDocument{%
  }
\setcopyright{none}
\usepackage{amsmath}
\usepackage{placeins}
\usepackage{graphicx} 		
\usepackage{subfigure}		
\usepackage{colortbl}
\usepackage{array}
\usepackage{multirow,makecell}
\usepackage{caption}
\usepackage{subcaption}
\usepackage{appendix}
\usepackage{booktabs}
\usepackage{svg}
\usepackage{longtable}
\usepackage{lscape}
\usepackage{pdflscape}
\usepackage[normalem]{ulem}
\usepackage{geometry}
\usepackage{xcolor}

\begin{document}


\title{Power Echoes: Investigating Moderation Biases in Online Power-Asymmetric Conflicts}



\author{Yaqiong Li}
\email{liyq22@m.fudan.edu.cn}
\affiliation{%
  \institution{Fudan University}
  \city{Shanghai}
  \country{China}
}

\author{Peng Zhang}
\authornote{Corresponding authors.}
\email{zhangpeng_@fudan.edu.cn}
\affiliation{%
  \institution{Fudan University}
  \city{Shanghai}
  \country{China}
}

\author{Peixu Hou}
\authornotemark[1]
\email{peixu.hou@meituan.com}
\affiliation{%
  \institution{Meituan}
  \city{Shanghai}
  \country{China}
}

\author{Kainan Tu}
\email{kntu25@m.fudan.edu.cn}
\affiliation{%
  \institution{Fudan University}
  \city{Shanghai}
  \country{China}
}

\author{Guangping Zhang}
\email{gpzhang20@fudan.edu.cn}
\affiliation{%
  \institution{Fudan University}
  \city{Shanghai}
  \country{China}
}

\author{Shan Qu}
\email{qushan@meituan.com}
\affiliation{%
  \institution{Meituan}
  \city{Shanghai}
  \country{China}
}

\author{Wenshi Chen}
\email{wenshi.chen@meituan.com}
\affiliation{%
  \institution{Meituan}
  \city{Shanghai}
  \country{China}
}

\author{Yan Chen}
\email{ych@vt.edu}
\affiliation{%
  \institution{Virginia Tech}
  \city{Virginia}
  \country{USA}
}

\author{Ning Gu}
\email{ninggu@fudan.edu.cn}
\affiliation{%
  \institution{Fudan University}
  \city{Shanghai}
  \country{China}
}

\author{Tun Lu}
\authornotemark[1]
\email{lutun@fudan.edu.cn}
\affiliation{%
  \institution{Fudan University}
  \city{Shanghai}
  \country{China}
}

\renewcommand{\shortauthors}{Yaqiong Li et al.}

\begin{abstract}

Online power-asymmetric conflicts are prevalent, and most platforms rely on human moderators to conduct moderation currently. Previous studies have been continuously focusing on investigating human moderation biases in different scenarios, while moderation biases under power-asymmetric conflicts remain unexplored. Therefore, we aim to investigate the types of power-related biases human moderators exhibit in power-asymmetric conflict moderation (RQ1) and further explore the influence of AI's suggestions on these biases (RQ2). For this goal, we conducted a mixed design experiment with 50 participants by leveraging the real conflicts between consumers and merchants as a scenario. Results suggest several biases towards supporting the powerful party within these two moderation modes. AI assistance alleviates most biases of human moderation, but also amplifies a few. Based on these results, we propose several insights into future research on human moderation and human-AI collaborative moderation systems for power-asymmetric conflicts.

\end{abstract}



\maketitle

\section{Introduction}
Various digital platforms like Twitter and Yelp have been important medium to support social interactions and activities. As the user base expands on these platforms, conflicts among different roles have become prevalent, posing challenges to platform's regulation. For example, conflicts between consumers and merchants over product quality or service problems occur frequently on e-commerce platforms like Yelp. According to the Global Digital Trust \& Safety Index \cite{Report_1}, millions of such conflicts were reported worldwide in the third quarter of 2023. In China, the Consumers Association reported over 780,000 consumer complaints in the first half of 2024, with a year-on-year increase of 27.21\% \cite{Yingchao_2}. Similar conflicts also arise on digital labor platforms such as Amazon Mechanical Turk and Upwork, where workers and requesters frequently disagree over issues like compensation and task quality \cite{Benjamin_3, Difallah_14, Martin_15}. 

These conflicts between different roles generally have a prominent structural characteristic, i.e., they often occur within relationships with power asymmetry \cite{Kolovson_4, Vallet_5, Passmore_16}. One party usually possesses more expertise and resources (e.g., merchants or employers) while the other party lacks conflict resolution experiences and relies heavily on platform's coordination (e.g., consumers or workers). Taking the online shopping scenario as an example, when consumers complain about the quality of the goods and give negative reviews, merchants may invoke the platform's ``malicious evaluation handling rules'' to file a complaint and present fine-grained explanations \cite{Hessamedin_18}. In contrast, consumers' negotiation skills, rule interpretation ability, and appeal channels are generally limited, and they struggle to provide more evidence, making their reviews ultimately determined to be inappropriate. 

Many platforms have introduced moderation strategies to handle the power-asymmetric conflicts. Different from traditional content moderation like toxicity moderation \cite{Li_49, Li_124} and privacy moderation \cite{Liu_51}, power-asymmetric conflict moderation requires making a judgment on ``who is more credible and who should be supported'' in the context of the power asymmetry and the ambiguous statements of both parties, aggravating the difficulty of moderating. Therefore, most platforms currently rely on human volunteers or crowd moderators to conduct power-asymmetric conflict moderation. However, existing research has shown that in conflict scenarios involving subjective judgments, human moderators tend to exhibit different moderation biases, such as bias in stance or language style. For example, through a field study of social content moderation, Roberts et al. found that human moderators struggle to remain consistently neutral, as their views tend to be shaped by policy ambiguity, task pressures, or personal beliefs. This study also points out that these factors can lead to moderators' biased judgments against certain groups, especially those users whose language expressions are relatively direct, vernacular, or lack strategic words—often described as low-context expressers \cite{Gerrard_6}. In addition, Seering et al. reported that Reddit moderators tend to reinforce existing power structures and dominant community values when moderating user conflicts, leading to marginalized voices being overlooked or removed \cite{Joseph_7}. Moreover, the ``underdog effect'' in psychology suggests that in conflict judgments, human moderators may sympathize with weaker parties and favor them, while they may also simply default to attributing responsibility to weaker parties when their statements are ambiguous \cite{Glick_8}. These findings imply that in power-asymmetric conflict moderation practice, judgments of human moderators can be not only influenced by personal stance and preferences, but also further amplified by the social power cues of the powerful party (e.g., language style, rule citation, and expression confidence), affecting the impartiality of the moderation process. Therefore, we pose the first research question (\textbf{RQ1}): \textbf{What types of power-related biases do human moderators exhibit when moderating power-asymmetric conflicts?}

With the advancement of AI tools powered by machine learning and deep learning, especially Large Language Models (LLMs), AI-assisted moderation has increasingly become more and more prevalent on different platforms \cite{Gaole_9}. They are widely utilized for generating explanations, offering decision-making suggestions, and determining risk levels, significantly enhancing the efficiency of content moderation \cite{Amit_54, Li_49, Mishra_52}. However, existing research indicates that the extent to which human moderators accept AI-generated suggestions is also influenced by people’s perceptions that the suggestion is generated by AI \cite{Hwang_12, Rosbach_72, Boonprakong_75}. This reveals an algorithmic compliance effect, where users can pose different attitudes towards ``\textit{This is the judgment result from GPT-4 large model.}'' Therefore, we further propose the second research question (\textbf{RQ2}): \textbf{How will AI-generated suggestions influence human moderators' power-related biases in power-asymmetric conflict moderation? Will the biases be amplified or mitigated?}

To address these two research questions, we focus on two judgment modes, i.e., human moderation and human–AI moderation, and systematically explore the corresponding biases in power-asymmetric conflict moderation. Based on the six power types defined in the Bases of Social Power theory \cite{AlKilani_21, French_20}, we take the conflicts between consumers and merchants as a representative power-asymmetric scenario, and collect real data from Dianping platform (a Chinese leading local life platform) for moderation bias investigation. Through coding, we extract power manifestations matched with the online conflict context and then introduce a taxonomy of potential power-related biases in power-asymmetric conflict moderation, such as legitimate claim bias, punishment threat bias, and compensation bias. Subsequently, we adopt a mixed design including between-subjects and within-subjects, and develop a web-based judgment program called ``I Support'' to conduct the experiment. A total of 50 participants were recruited and randomly assigned to the human moderation group or the human–AI moderation group. For \textbf{RQ1}, participants in the human moderation group read power-asymmetric conflict samples independently without external suggestions. Each conflict sample contains a specific power manifestation (e.g., the legitimate claim or compensation), and participants are required to judge which party they would like to support using a 5-point Likert scale. In our study, ``support a party'' means that the moderator considers the party’s statement more reasonable or credible. By comparing their decisions in different situations, we can identify which power manifestations are more likely to trigger biased judgments, thereby uncovering biases inherent in human moderation. For \textbf{RQ2}, for the same conflict samples, participants in the human–AI moderation group are provided with additional AI-generated suggestions. To control for potential variations in the outputs of different LLMs, we employ a Wizard-of-Oz design. This approach allows us to examine how participants' perceptions of AI-generated suggestions shape their moderation decisions. Specifically, we craft high-quality suggestions derived from crowdsourced data from real moderators  in advance and inform participants that the suggestions are generated by ``AI''. By comparing the judgments of the two groups under different power manifestations, we uncover whether AI assistance would amplify or mitigate different biases. Several findings have emerged from our quantitative and qualitative analyses. For RQ1, human moderation shows five power-related biases towards supporting the powerful party. For RQ2, these biases persist in human-AI moderation. Although AI assistance alleviates most biases in human moderation (four alleviated and one eliminated), it also amplifies a few (one introduced and another amplified). Besides, the perspective for unsupporting the opposite party in AI-generated suggestions can motivate moderators to support the weaker party. These findings provide empirical evidence for improving the moderation process and helping to alleviate moderation biases. Overall, this study makes the following contributions.
\begin{itemize}
    \item To the best of our knowledge, this is the first study on investigating the biases of human moderation and human–AI moderation under power-asymmetric conflicts.
    \item Based on real data, we develop a theory-driven taxonomy that summarizes potential power-related biases with corresponding power manifestations in power-asymmetric conflict moderation. Through an experiment with 50 participants, we uncover the biases in human moderation and the impact of AI assistance.
    \item We propose several insights into the design of power-asymmetric conflict moderation mechanisms and AI-assisted moderation systems. 
\end{itemize}

The rest of this paper is organized as follows. Section 2 reviews the related work. Section 3 describes power manifestations. Sections 4 and 5 present the experimental method and results. Sections 6  and 7 discuss implications and limitations. Finally, Section 8 concludes the paper.

\section{Related Work}
In this section, we review the theoretical basis of power asymmetry and its manifestation on digital platforms. We also draw on HCI research on conflict and moderation bias.

\subsection{Power-Asymmetric Conflicts}
Power asymmetry (or power inequality) refers to ``\textit{a state in which differences in status exist between individuals and groups of individuals, and these differences result in differential ability to take action or cause action to be taken}'' \cite{Passmore_16, Fousiani_17}. In social psychology and organizational research, power asymmetry is manifested in differences regarding resources and decision-making authority. Early studies on social relations and resource dependence provide a foundation for understanding power. Dahl proposed the concept of power as a relation among people, indicating the extent to which one party can influence the behavior of another party \cite{Dahl_108}. Emerson's power dependence model further pointed out that power is not an individual attribute but a function of mutual dependence. If one party relies on the other for critical resources, the party being relied upon holds greater power. This dependence is jointly shaped by the importance and substitutability of the resources \cite{Emerson_109}. To describe power systematically, researchers have also attempted to classify power sources in social interactions. The ``Bases of Social Power'' theory is widely recognized and categorizes power sources into six types: legitimate power (the authority conferred by positions or roles), reward power (the capacity to influence others' behavior by providing rewards), coercive power (the capacity to influence others' behavior by imposing punishments), expert power (the influence based on professional knowledge or skills), referent power (the following effect resulting from respect or admiration for others), and informational power (the ability to influence others’ behavior by controlling, disseminating or manipulating the information) \cite{Fousiani_17, French_20, AlKilani_21}. The theory offers a universal framework for understanding power asymmetry in different scenarios.

With the rapid growth of Internet platforms, people's social activities are increasingly shifting from offline to online. Research in social psychology and HCI fields has also begun extending power theories to digital contexts \cite{Panpan_102, Ye_103, Hou_104}. Prior work shows that digital technologies have changed the way users obtain and process information through approaches like interface design, personalized recommendation, content ranking and filtering \cite{Kling_110, Schnabel_111, Munoz_112}, thereby influencing what users can see, how much they can see, and in what order they can see it. These changes may cause an imbalance in the distribution of information, reconstructing or even exacerbating the resource disparity among users and giving rise to new forms of power asymmetry in different online scenarios \cite{Kling_110, Markus_113}. In online communities, newcomers generally rely on older users' content and guidance; thus, old users have strong expert and informational power due to their knowledge and expertise \cite{Saxena_114}. In social media, celebrities' posts usually can gain higher exposure due to their social appeal, while the content of ordinary users struggles to achieve comparable popularity \cite{Chou_115, Xiao_116}. These resource gaps can limit the action space of the weak party in interactions, or even lead to confrontational or conflict behaviors, i.e., power-asymmetric conflicts. Although power asymmetry research originates in offline interaction scenarios, power-asymmetric conflicts in the online environment still essentially stem from the inequality of information, influence, and other resources among users. However, digital environments introduce more resources (e.g., posts, followers, account reputation), all of which may increase the probability of resource asymmetry. Therefore, the manifestations reflected in existing power theories may necessitate expansion when being transferred into digital contexts.

E-commerce and digital labor platforms, whose resource structure has undergone significant changes compared to offline scenarios, are representative digital contexts easily occurring power-asymmetric conflicts \cite{Janice_117, Ikhsan_118, Martin_15, Panpan_102}. On e-commerce platforms, merchants often control critical resources like product information, pricing strategies, and after-sales services \cite{Hessamedin_18, Nyaga_19}, which places them in a position of information superiority in transactions. Conversely, consumers are highly reliant on the information provided by merchants via the platform, giving rise to information asymmetry. This imbalance implies that when issues arise, consumers generally encounter difficulties in seeking effective channels for safeguarding their rights. Constrained by the rules of merchants or platforms, they struggle to express their needs, prompting them to adopt alternative strategies such as public reviews to assert their rights \cite{Janice_117}. Similar phenomena are also observed in the digital labor platforms \cite{Difallah_14, Martin_15}. Workers are usually at a disadvantage in terms of information access and resources. Lacking sufficient negotiation skills, they are prone to accept low wages or unequal task requirements. By contrast, requesters can gain stronger control through their familiarity with platform regulations, the power to issue tasks, and their influence on the rating system. Consequently, in situations of overdue payment or task-related disputes, workers generally find it hard to seek compensation. They also face potential risks such as lower ratings or even account suspension \cite{Martin_15}. This power asymmetry between different parties can not only impact personal rights but also trigger online conflicts. Moreover, issues such as the lack of transparency in conflict resolution processes or biases can further exacerbate trust crises among parties \cite{Munoz_112}.

\subsection{Conflict Analysis and Resolution Studies in HCI}
Conflict analysis and resolution is an important research topic in the field of Human–Computer Interaction (HCI) \cite{Easterbrook_22}, covering scenarios such as team collaboration \cite{Huang_29, Chen_36}, social interaction \cite{Kittur_27, Levy_28}, virtual reality \cite{Jagannath_24, Spivack_25}, and crowdsourcing \cite{Brachman_26}. Previous studies have explored reasons for different conflicts, specific manifestations in different environments, and impacts on interaction quality, user experience, and collaboration efficiency. In temporary design teams, conflicts are often triggered by inconsistent goals, divergent creative directions, and unequal task allocation \cite{Chen_36}. In social interactions, user attributes, conversation history, and emotional fluctuations are suggested to be related to the degree of group conflict \cite{Levy_28}. In online peer review process, feedback that only contains negative conclusions or emotional reactions without actionable measures can cause relational conflicts among contributors, reducing some members' willingness to continue participating in the community \cite{Huang_29}. An analysis of conflict discussions on Twitter has revealed that conflicts can not only decrease the number of participants, but also weaken users' enthusiasm and collaboration efficiency \cite{Canute_61}.

HCI researchers have further examined conflict resolution strategies in different contexts. These strategies can be categorized into three types based on the practitioners: human-led \cite{Jagannath_24, Spivack_25}, AI-led \cite{Shaikh_33}, and human–AI collaboration \cite{Govers_34}. Human-led resolution approaches rely on the expertise, experience, and social influence of human moderators. In virtual gaming environments, caring adults and near-peer mentors play a crucial role in defusing disputes. They achieve this through providing explanatory feedback, reminding participants of rules, and suggesting problem-solving strategies \cite{Jagannath_24}. In team collaboration, researchers have proposed various methods to foster a positive team atmosphere, such as introducing ``team dates'' to enhance communication \cite{Curseu_30}, encouraging silent members to contribute \cite{Kim_31}, and promoting balanced participation among members \cite{Tausczik_32}. For AI-led conflict resolution, techniques like machine learning or LLMs are leveraged directly to mediate conflicts, which is an emerging and promising approach for conflict moderation. Omar et al. developed an interactive LLM-powered system called \textit{Rehearsal} that helps users practice skills for handling interpersonal conflicts through role playing and real-time feedback \cite{Shaikh_33}. Finally, human–AI collaboration resolution combines human understanding with the computational advantages of AI. In online debates on social platforms, researchers have employed prompt engineering techniques to incorporate multiple conflict resolution strategies (e.g., competition, avoidance, and compromise) into LLM-generated responses. This method aims to assist users in reducing hostile language and alleviating polarization \cite{Govers_34}. Overall, these studies have improved the understanding of conflict causes and resolution mechanisms. However, most studies assume equal status between the conflict parties, and there is insufficient exploration of power-asymmetric conflicts.

\subsection{Fairness and Bias in Moderation}
Content moderation is defined as ``\textit{the governance mechanisms that structure participation in a community to facilitate cooperation and prevent abuse}'' \cite{Grimmelmann_40}. Moderation practices often face trade-offs among different behavioral norms, communicative styles, and value orientations, and the final decisions depend on several factors like specific contexts and platform rules \cite{Jiang_39}. Current content moderation mainly includes two approaches: human moderation and AI-assisted moderation. 

As the primary method, human moderation refers to employing human moderators to identify inappropriate content such as hate speech or harassment \cite{Jhaver_44, Kiesler_45}. However, prior research shows that human moderation is not always neutral. Moderators can be influenced by factors such as personal characteristics, contextual ambiguity, and cognitive or workload constraints \cite{Seering_41, Tim_42}. First, identity-related factors can influence how moderators interpret user behavior. For example, moderators' decisions can be affected by personal characteristics, including demographics, social status, political stance, cultural background, etc \cite{Joseph_38}. For example, moderators tend to review female streamers (users who engage in live streaming) than male streamers on Twitter \cite{Andrew_46}. Posts from political conservatives, transgender users, and Black users are prone to being moderated and removed \cite{Haimson_37}. Second, perceptual biases can emerge when moderators face ambiguous narratives or policy uncertainty. Field studies show that moderators struggle to maintain neutrality when policies are vague or when task pressure is high, leading to strict moderation towards users whose communication is more direct, vernacular, or lacks strategic linguistic words \cite{Gerrard_6, Glick_8}. Besides, heuristic biases can arise under heavy workloads. As online communities grow in content size, human moderators often encounter extensive work and high workloads, which impair their attention and judgment \cite{Wohn_43, Ford_47}. Moderators may overly focus on certain types of inappropriate content, while exhibiting reduced sensitivity to others.

Biases in algorithmic and AI-assisted moderation systems have also drawn growing attention. With the rapid development of techniques, the use of AI in content moderation has become increasingly prevalent, especially in the detection of toxic or harmful content \cite{Lai_50, Shahid_58, Fan_74}. By leveraging machine learning \cite{Ibrohim_53}, deep learning \cite{Liu_51, Amit_54}, and LLMs \cite{Ye_10, Chen_11, Li_49, Mishra_52}, AI-assisted moderation systems can efficiently detect online toxic or harmful content. 
A notable example is Google’s Perspective API, which is widely used on platforms like Reddit and Faceit. Trained on large corpora, this API can identify toxic content and assign a toxicity score \cite{Perspective_55}. However, research indicates that its judgment is sometimes biased. The same content often receives a higher toxicity score in German compared to other languages \cite{Nogara_48}. This disparity highlights a broader issue that models trained on large datasets inherit the biases embedded in those datasets. Empirical studies show that LLMs frequently exhibit systematic biases in judgment tasks, showing sensitivity to various factors like input length and input order \cite{Ye_10, Chen_11}. Moreover, some new forms of biases have emerged in human–AI moderation \cite{Alon_56, Prabhudesai_73}. Prior research has found that human moderators tend to over-rely on algorithmic suggestions, i.e., a phenomenon known as the algorithmic compliance effect, which weakens independent critical thinking and makes judgments highly consistent with AI outputs \cite{Alon_56}. At the community level, members may selectively adopt AI-generated suggestions \cite{Kou_57, Schoeffer_66}. They tend to accept content that aligns with personal expectations or group identity, while questioning or rejecting those that contradict their views \cite{Kou_57}.

Overall, conflict analysis and resolution is an important research topic in HCI, particularly regarding the strengths and limitations of human–AI moderation. As a new form of online conflict, power-asymmetric conflict is characterized by differences in user status, informational imbalance, and an increasing trend of moderation bias. However, existing research still lacks a systematic investigation of biases that arise in such power-asymmetric moderation tasks. To fill this gap, our study examines the potential power-related biases in human moderation and further introduces AI assistance to explore whether it mitigates or amplifies these biases. Through this, we aim to provide new insights for the design of moderation strategies and human–AI moderation systems for power-asymmetric conflicts. 

\section{Power Manifestations in Online Power-asymmetric Conflicts}
This paper aims to explore the power-related biases exhibited by moderators in power-asymmetric conflict moderation. We first review the power types in social interactions and derive the corresponding potential bias categories according to existing theories and empirical studies. Specifically, we draw on the ``Bases of Social Power'' theory proposed by social psychologists French \& Raven, which divides social power into six categories: legitimate power, coercive power, reward power, expert power, referent power, and informational power \cite{French_20, AlKilani_21}. This framework has been widely applied in offline interpersonal interaction, organizational communication, and negotiation research \cite{Fousiani_17, Petress_59, Ten_60}, and has also been used to explain power dynamics in online interactions in recent years \cite{Panpan_102, Ye_103, Hou_104}. These applications provide a theoretical lens for characterizing power relations and understanding power asymmetry in online conflicts. Therefore, this study takes the ``Bases of Social Power'' theory as a reference and conducts an empirical analysis to systematically investigate the manifestations of power types in online power-asymmetric conflicts by using online consumer-merchant conflicts as the research scenario. This can offer a basis for our subsequent bias analysis.

\begin{table*}[ht]
    \centering
    \small 
    \setlength{\tabcolsep}{1pt}
    \caption{Ten forms of power manifestation in online power-asymmetric conflicts.}
    \begin{tabular}{>{\centering\arraybackslash}m{0.12\linewidth}
                    >{\raggedright\arraybackslash}m{0.26\linewidth}
                    >{\centering\arraybackslash}m{0.15\linewidth}
                    >{\raggedright\arraybackslash}m{0.4\linewidth}}
    \toprule
    \textbf{Power Type} & \multicolumn{1}{c}{\textbf{Description of Power Type}} & \textbf{Power Manifestation} & \multicolumn{1}{c}{\textbf{Description of Power Manifestation}} \\
    \midrule
    
    \multirow{3}{*}{\parbox[c]{1.8cm}{\centering Legitimacy power}}
    & \multirow{3}{=}{It comes from an elected, selected, or appointed position of authority and may be underpinned by social norms or authority citation} 
    & Legitimate claim 
    & Strengthening one's position by citing rules (e.g., \textit{"This is the platform's rule"}) \\
    \cmidrule(lr){3-4}
    & & Authority citation 
    & Supporting one party by referring to external institutions or sources (e.g., \textit{"please visit http : / /..."}) \\
    \midrule
    
    Coercive power &
    It uses the threat of force to gain compliance from another &
    Punishment threat &
    Forcing the other party to give in by threats, complaints, or negative consequences (e.g., \textit{“I'm going to complain about you”}) \\
    \midrule
    
    Reward power &
    It is based on the right to offer or deny tangible, social, emotional, or spiritual rewards &
    Compensation &
    Offering benefits to win support (e.g., \textit{“You can get a full refund”}) \\
    \midrule
    
    Expert power &
    It is based on knowledge, experience, and specialized skills &
    Expert knowledge &
    Supporting one party by providing professional knowledge (e.g., \textit{“The quality meets the required standard”}) \\
    \midrule
    
    Referent power &
    It is based on respect or admiration leading to identification or emulation &
    Group preference &
    Supporting one party by referring to group consensus (e.g., \textit{“Everyone says the service is bad”}) \\
    \midrule

    \multirow{7}{*}{\parbox[c]{1.8cm}{\centering Informational power}} &
    \multirow{7}{=}{It is based on controlling information needed by others to reach important goals} &
    Statement order &
    Influencing perception by speaking first or last \\
    \cline{3-4}
    & & Expression tone &
    Strengthening one's position through emotional language (e.g., \textit{“It is so disgusting”}) \\
    \cline{3-4}
    & & Choice trap &
    Highlighting rationality through comparison (e.g., \textit{“Other stores offer better service”}) \\
    \cline{3-4}
    & & Length difference &
    Influencing perception through differences in response length \\
    
    \bottomrule
    \end{tabular}

    \label{tab:type_mapping}
\end{table*}

\textit{\textbf{Data Collection.}} We choose the Dianping platform\footnote{https://www.dianping.com} as the research platform for three reasons: 1) Dianping is a leading local life service platform in China, covering fields such as dining, hotels, entertainment, and tourism, with over 250 million users. Its functions are similar to those of Yelp and TripAdvisor; 2) There have been a large number of disputes and conflicts between consumers and merchants in the comment area of Dianping, which exhibit the characteristics of power asymmetry such as information asymmetry and differences in role status; 3) The platform offers a ``Public Review'' function, which is a typical crowdsourcing moderation method. It publicly presents conflict samples and allows different users to vote, and also displays the voting results and the detailed reasons, providing a rich corpus for our research. For data collection, we obtain approximately 60,000 consumer-merchant conflict samples since April 2025, covering 17 topics. Subsequently, we select the top five topics by quantity (accounting for 92.7\% of total samples), namely ``\textit{Food}'', ``\textit{Beauty}'', ``\textit{Entertainment}'', ``\textit{Sports}'', and ``\textit{Hotels}''. For each topic, we select 20 samples: 10 where the crowdsourced label the considers consumer reviews reasonable and 10 where it considers merchant responses reasonable. The crowdsourced label derives from moderators’ judgments through the platform's ``Public Review'' function. Each sample contains consumer reviews, merchant responses, the crowdsourced label, and corresponding reasons from moderators. Finally, our dataset comprises 100 samples.

\textit{\textbf{Data Analysis and Results.} }
During the data analysis stage, based on the ``Bases of Social Power'' theory, we examine the specific manifestations of six power types in our online conflict corpora. We adopt a qualitative analysis method to code and categorize the conflict corpora. Specifically, three authors with backgrounds in human-computer interaction, social psychology, and linguistics collaborate to verify the manifestations of the six power types in the corpora. For instance, the statement ``\textit{This is the platform's rule}'' in consumer reviews reflects the legitimate power, while the merchant's response ``\textit{If you are not satisfied, you can get a full refund}'' belongs to the reward power. Through this process, we map the samples to the corresponding power types. Finally, all conflict samples can be covered by the six power types, and no new types beyond the framework have emerged. This verifies the applicability of this theory in online power-asymmetric conflicts.

We conduct a qualitative analysis and refine the manifestations of the six power types in our conflict corpora \cite{Wicks_105, Miles_106, Zhang_13}. Three authors first independently code a pilot subset containing 25 samples (each topic contains 5 samples) using open coding, without presupposing subcategories. During this stage, the authors mark any textual features that indicate a potential expression of six power types. For example, ``\textit{waiting for being punished}'' in consumer reviews is labeled as ``punishment threat'', and ``\textit{need to discount}'' or ``\textit{free meal}'' in merchant responses are labeled as ``compensation''. After completing the initial open coding, the authors conduct multiple rounds of discussion to consolidate codes, resolve disagreements, and develop a shared codebook. This process results in 38 initial power manifestations. Subsequently, we conduct axial coding to merge semantically similar codes and map them to six power types. For example, ``\textit{compensation}'', ``\textit{need to discount}'', and ``\textit{free meal}'' are merged into the ``compensation'' manifestation (corresponding to the reward power); ``\textit{longer text}'' and ``\textit{shorter text}'' are merged into ``length difference'' (corresponding to the informational power). Saturation is reached when no new power manifestations emerge in two consecutive rounds of coding. To ensure the reliability, three authors select 25 samples randomly from the remaining 75 and code them using the refined codebook. We perform a consistency check using Fleiss’s Kappa and obtain a score of 0.76, which represents substantial agreement \cite{McDonald_107}. The remaining samples are coded by two authors, and disagreements are resolved through discussion with the third author. Ultimately, we extract ten forms of power manifestation, including legitimate claim, authority citation, punishment threat, compensation, expert knowledge, group preference, statement order, expression tone, choice trap, and length difference (see Table \ref{tab:type_mapping} for details).

Each power manifestation may potentially influence the moderator's judgment in power-asymmetric conflicts. Taking the legitimate claim bias as an example. When citing rules, one party's statements seem to be more reliable, making moderators more inclined to support the party. Similarly, the group preference represents that one party uses group consensus to support the viewpoint, which may prompt moderators to accept the majority opinion rather than making decisions based on independent judgment. According to this analysis, we can obtain 10 power-related bias types corresponding to the 10 power manifestations, including legitimate claim bias, authority citation bias, punishment threat bias, compensation bias, expert knowledge bias, group preference bias, statement order bias, expression tone bias, choice trap bias, and length difference bias. This taxonomy provides a foundation for subsequent experiment design to examine moderators' biases under different power types.

\begin{figure*}[ht]
  \small
  \centering
  \includegraphics[width=0.9\linewidth]{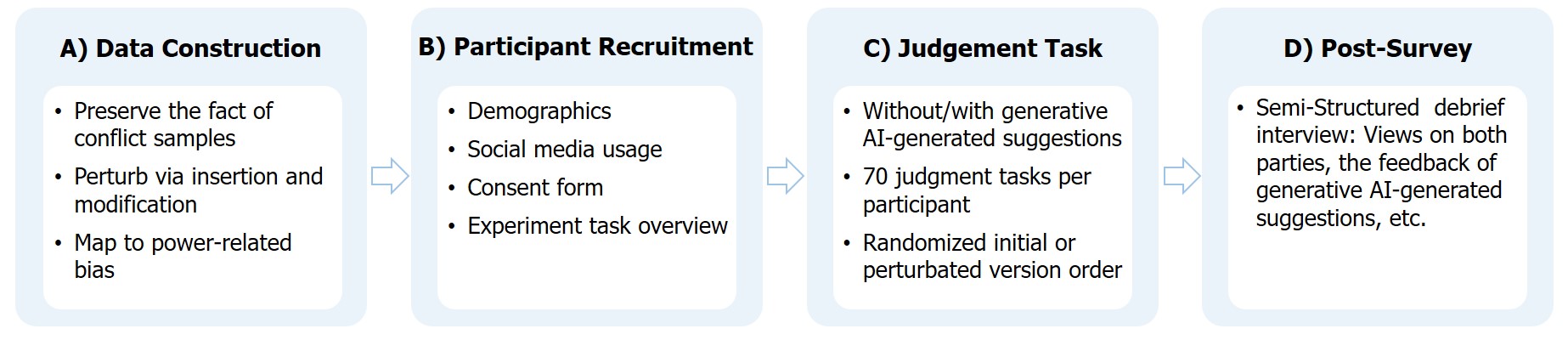}
  \caption{The flow of experimental design.}
  \Description{The figure illustrates the flow of experimental design, which includes four components: data construction, participant recruitment, the judgment task, and the post-survey.}
\label{fig:flow}
\end{figure*}

\section{Experimental Design}
Based on the power-asymmetric conflict taxonomy, we employ an intervention method to explore the power-related biases of moderators \cite{Quinn_63}. The overall flow is shown in Figure \ref{fig:flow}. The intervention method introduces bias-related perturbations (such as adjusting the order of information presentation or adding noise content) in experiments and compares the judgment differences between the original situation and the perturbed conditions, thus revealing potential bias patterns. This method has been widely used in fields such as human-computer interaction and natural language processing \cite{Krsek_64, Yang_65, Chen_11, Divyansh_67}. Specifically, based on the power-related bias taxonomy constructed in the previous section, we initially prepare experimental data by adding perturbations associated with each bias to the real conflict corpus. Then we adopt a mixed experimental design with both between-subjects and within-subjects variables. The between-subjects variable distinguishes between the human and human-AI moderation modes, while the within-subjects variable is that each participant randomly reads the initial or perturbed version of the conflict samples. After participants complete the experiment, we employ semi-structured interviews to collect participants' views on different conflict examples and their reflections on AI-generated suggestions. Combining the judgment results and interview data, we analyze the differences between moderation modes under various perturbations. The following sections will present details of experimental data construction, participant recruitment, and the experimental procedure.

\subsection{Data Construction}
This section details the method for constructing experimental data based on the taxonomy. For each bias, we perturb the real consumer-merchant conflict samples to generate experimental materials. The perturbation process adheres to three principles: 1) The core facts of the conflict are preserved; 2) Only the perturbation related to one power manifestation is inserted or modified within the statement of one party; 3) Each sample contains only one power manifestation to avoid cross-influences. Most manifestation perturbations are completed through insertion. For example, adding the statement like ``\textit{This is the platform's rule}'' (legitimate claim) to the merchant's response or ``\textit{Everyone says the service is bad}'' (group preference) to the consumer's review. Perturbations for some manifestations such as statement order and expression tone are achieved through modification. Based on the 100 initial samples (see Data Collection in Section 3), we construct data in terms of 10 perturbations. 9 perturbations (except the statement order) are constructed in two directions: adding perturbations to the consumer reviews while keeping the merchant responses unchanged, and vice versa. This approach yields paired samples, which help us analyze from two parties and avoid potential bias. Finally, 2,000 samples are obtained (100 initial and 1,900 perturbed samples). Table \ref{tab:conflict_samples} presents an initial sample and the corresponding perturbed versions. We will publicly release the initial and perturbed data in the future, thus facilitating further investigations of the judgment mechanism in power-asymmetric conflict moderation.

\begin{table*}[ht]
    \centering
    \small
    \setlength{\tabcolsep}{1pt}
    \caption{The initial and corresponding perturbed conflict samples (all samples are translated from Chinese for understanding).}
    \label{tab:conflict_samples}
    
    \begin{tabular}{>{\centering\arraybackslash}m{0.13\linewidth}
                    >{\raggedright\arraybackslash}m{0.25\linewidth}
                    >{\raggedright\arraybackslash}m{0.55\linewidth}}
        \toprule
        \multicolumn{3}{c}{\textbf{Initial Conflict Sample}} \\
        \midrule
        \multicolumn{3}{p{0.95\linewidth}}{\textbf{Consumer's review}: \textit{So this new ``internet-famous'' bakery just opened, and guess what? They're already doing that whole ``only a few people inside at a time'' thing. On top of that, you have to pay separately, which is super annoying. Honestly, the bread isn't even that great, definitely not worth the price. And now it looks like they're even paying people to write fake positive reviews.}} \\
        \multicolumn{3}{p{0.95\linewidth}}{\textbf{Merchant's response}: \textit{During our grand opening, we offered special promotions and the turnout was much bigger than expected. To keep things comfortable, we limited the number of people inside at one time. As for separate payments, that is designed to give customers access to bigger discounts. We never thought it would be taken as ``hype marketing''.}} \\
    \end{tabular}
    
    \begin{tabular}{>{\centering\arraybackslash}m{0.13\linewidth}
                    >{\raggedright\arraybackslash}m{0.25\linewidth}
                    >{\raggedright\arraybackslash}m{0.55\linewidth}}
          \midrule
        \multicolumn{3}{c}{\textbf{Perturbed Conflict Sample}} \\
        \midrule
        \textbf{Power-related Bias} & \textbf{Perturbation Manipulation} & \multicolumn{1}{c}{\textbf{Example}}\\
        \midrule
        
        Legitimate claim bias & Insert content related to platform rules in one party's statement &
        \textbf{Consumer's review}: \textit{So this new ``internet-famous'' bakery just opened... the price, \textbf{which goes against the Merchant Pricing Guidelines}. And now it looks...} \newline
        \textbf{Merchant's response}: \textit{During our grand opening, we offered special...} \\
        \midrule
        
        Authority citation bias & Insert empty links with no content in one party's statement  &
        \textbf{Consumer's review}: \textit{So this new ``internet-famous'' bakery just opened...} \newline
        \textbf{Merchant's response}: \textit{During our grand opening... inside at one time. \textbf{You can see the rules here: http://...} As for separate payments...} \\
        \midrule
        
        Punishment threat bias & Insert content about the possible penalties or negative consequences that the other party may face &
        \textbf{Consumer's review}: \textit{So this new ``internet-famous'' bakery just opened.... positive reviews. \textbf{I'm so calling this place out online!}} \newline
        \textbf{Merchant's response}: \textit{During our grand opening, we offered special...} \\
        \midrule
        
        Compensation bias & Insert a commitment to compensation or rewards in one party's statement &
        \textbf{Consumer's review}: \textit{So this new ``internet-famous'' bakery just opened...} \newline
        \textbf{Merchant's response}: \textit{During our grand opening... inside at one time. \textbf{We also gave out free snacks and little gifts to customers who couldn't get in.} As for separate payments...} \\
        \midrule
        
        Expert knowledge bias & Insert professional terms or technical language in one party's statement  &
        \textbf{Consumer's review}: \textit{So this new ``internet-famous'' bakery just opened...} \newline
        \textbf{Merchant's response}: \textit{During our grand opening... inside at one time. \textbf{We implemented dynamic adjustments based on real-time traffic and used a high-frequency diversion mechanism to optimize.} As for separate payments...} \\
        \midrule
        
        Group preference bias & Insert position content of the majority in one's party statement  &
        \textbf{Consumer's review}: \textit{So this new ``internet-famous'' bakery just opened... positive reviews. \textbf{A bunch of my friends have complained about it too, and we all feel like we got played!}} \newline
        \textbf{Merchant's response}: \textit{During our grand opening, we offered special...} \\
        \midrule
        
        Statement order bias & Switch the statement order of the consumer and the merchant  &
        \textbf{Merchant}: \textit{During our grand opening, we offered special...}  \newline
        \textbf{Consumer's review}: \textit{So this new ``internet-famous'' bakery just opened...}  \\
        \midrule
        
        Expression tone bias & Modify the language style of one party's statement  &
        \textbf{Consumer's review}: \textit{So this new ``internet-famous'' bakery just opened...} \newline
        \textbf{Merchant's response}: \textit{\textbf{We're sorry if the experience wasn't ideal}. During our grand opening... that is to give customers bigger discounts. \textbf{We understand the concerns about ``hype marketing'' and we'll take the feedback seriously as we work to improve the service experience.} } \\
        \midrule
        
        Choice trap bias & Insert content of alternative solutions in one party's statement  &
        \textbf{Consumer's review}: \textit{So this new ``internet-famous'' bakery just opened... positive reviews. \textbf{At least the bakery next door has better bread and fair prices.}} \newline
        \textbf{Merchant's response}: \textit{During our grand opening, we offered special...} \\
        \midrule
        
        Length difference bias & Insert vague or irrelevant content in one party's statement &
        \textbf{Consumer's review}: \textit{So this new ``internet-famous'' bakery just opened... positive reviews. \textbf{Oh, and the elevator was out of service when I went. Really bad luck!}} \newline
        \textbf{Merchant's response}: \textit{During our grand opening, we offered special...} \\
        \bottomrule
    \end{tabular}
\end{table*}
\subsection{Participant Recruitment and Demographics}
This study has obtained approval from the Institutional Review Board (IRB) of the first author's institution. All researchers have completed the human subject training. Participants need to be at least 18 years old, have fluent Chinese reading skills, and have basic experience using platforms such as Meituan, Dianping, and Ele.me. Participants were recruited publicly through online platforms and a questionnaire recruitment system. Before the experiment, we provide potential participants with a detailed description of experimental procedures, including guidelines of the online judgment program, the main task (70 judgment tasks), the semi-structured review interview (about 30 minutes), and the potential fatigue risks that may arise. When using the judgment program, referring to the daily task volume of Dianping's ``Public Review'' function, we suggest participants take a break after finishing 10 questions per day. All moderation tasks can be completed within one week. Each participant is required to sign an informed consent before participating in the study and can withdraw at any stage. We also provide instructions on how to delete personal annotation data and mental health support from social workers. During the experimental process, no identifiable personal information is collected, and all data is stored on a local platform. Researchers are not allowed to download data on personal devices. After completing the experiment, participants will receive compensation of 50 RMB.

Finally, 50 participants were recruited, with an average age of 26.18 years (SD = 4.45). 27 are male (54.0\%) and 23 are female (46.0\%). 20 participants have a bachelor's degree or lower (40.0\%), and 30 have a master's degree or higher (60.0\%). Moreover, 46 participants (92.0\%) have used platforms like Dianping for more than two years, and 2 (4.0\%) for more than one year.

\subsection{Experimental Task Design}
To explore potential power-related biases in human and human-AI moderation, we adopt a mixed design including between-subjects and within-subjects factors. The between-subjects factor is whether there is an AI-generated suggestion (human moderation group vs. human-AI moderation group), and the within-subjects factor is whether there is a perturbation (initial version vs. perturbed version). The experiment includes two phases.

\begin{itemize}
    \item \textbf{Phase 1: Online judgment task}. Participants read a series of conflict samples on the ``I Support'' program developed by us and determine which party is more reasonable and credible.
    \item \textbf{Phase 2: Semi-structured interview}. After all participants complete the tasks in Phase 1, we randomly select some participants from all groups to conduct a one-on-one interview to gain a deeper understanding of their thoughts during judgment.
\end{itemize}

\begin{figure*}[t]
  \small
  \centering
  \includegraphics[width=0.9\linewidth]{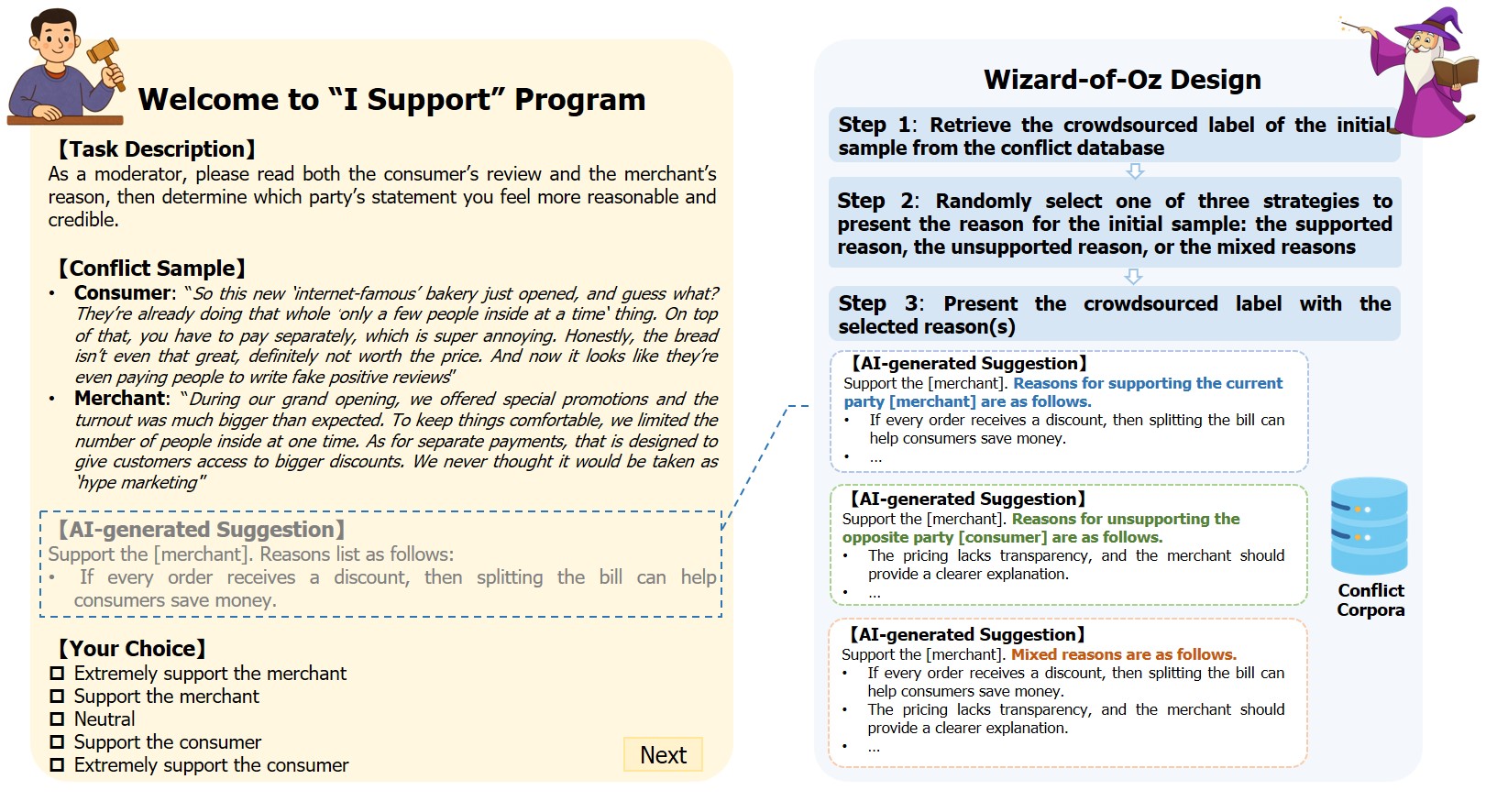}
  \caption{The design of ``I Support'' program (for presentation convenience, all descriptions are translated into English).}
  \Description{The figure illustrates the design of “I Support” program. The left panel shows the four components of the program: the task description, the conflict sample, the AI-generated suggestion (visible only to the human-AI moderation group), and the user choice. The right panel outlines the three steps of the Wizard-of-Oz procedure: retrieving the reference of the initial case from the conflict database, randomly selecting one of three strategies to present the reason, and presenting the reference label with the selected reason.}
\label{fig:program}
\end{figure*}

In the first phase, we develop a web-based judgment program called ``I Support'' (see Figure \ref{fig:program}, left), which consists of four parts: task description, conflict sample, AI-generated suggestion (visible only to the human-AI moderation group), and user choice. The presentation of conflict samples adopts the case-to-condition randomization method, which means that each participant would only be exposed to one version of a conflict sample (randomly selected from the initial version or 10 perturbed versions). This method can alleviate the influence of sequence effects and learning effects. For the human-AI moderation group, following previous research \cite{Gu_62, Maulsby_68}, we adopt a Wizard-of-Oz method (see Figure \ref{fig:program}, right), which leverages real researchers to provide high-quality suggestions while informing participants that the suggestions are generated by ``AI'' like LLMs. This design helps eliminate the impact of quality differences among various LLMs' outputs and enables us to isolate the effects of perceived AI-generated suggestions on moderation. To ensure the reliability of ``AI''-generated suggestions, we craft suggestions based on the crowdsourced label and corresponding reasons produced by crowd moderators (see Section 3). Existing research has shown that crowdsourced judgments are more reliable than individual judgments \cite{Ueda_69, Surowiecki_70}. Therefore, three authors review the corresponding reasons for 100 initial samples (50 samples are labeled as reasonable consumer statements and 50 as reasonable merchant statements by crowd moderators). Through multiple rounds of discussion, they reach a consensus on the high-quality reasons - those that demonstrate clear logic, reference the relevant context, and avoid emotional expressions. The selected reasons are then combined with the crowdsourced label as a ``AI''-generated suggestion (see Figure \ref{fig:program}, right). For the initial and perturbed versions of one conflict sample, AI-generated suggestions use the same crowdsourced label, but randomly present one of three reasons constructed from three perspectives (as described below). Besides, the user choice part uses a five-point Likert scale with options: extremely support the merchant, support the merchant, neutral, support the consumer, and extremely support the consumer. 

The reasons for a crowdsourced label can be expressed from three perspectives: just giving the reasons for supporting the current party, just giving the reasons for unsupporting the opposite party, and a mixed of the two kinds of reasons. When a crowd moderator supports the consumer in conflict moderation, she/he can describe her/his reasons for supporting the consumer, reasons for unsupporting the merchant, or both. As shown on the right of Figure \ref{fig:program}, these strategies could guide participants to think from different perspectives. For example, when participants see ``\textit{If every order receives a discount, then splitting the bill can help consumers save money}''(presenting the reasons for supporting the current party), they might interpret it as the merchant patiently explaining discount details to the consumer, thus being prone to accept the merchant's statements. When presented with ``\textit{The pricing lacks transparency, and the merchant should provide a clearer explanation}'' (presenting the reasons for unsupporting the opposite party), participants may pay attention to potential vague operations of the merchant. In samples presenting mixed reasons for both parties, participants can see the reasons from the above two perspectives.

In the second phase, we randomly select some participants from both groups and conduct a one-on-one semi-structured interview with their consent. The interview begins with recalling the judgment tasks completed by the participant in the first phase, including conflict samples he/she evaluated and corresponding choices he/she made. Then, we ask some questions about their judgment processes, covering aspects like crucial information or contextual factors that influenced moderation, the role of different power manifestations, and reasons for the trade-off among five options (ranging from extremely support the merchant to extremely support the consumer). For participants in the human–AI moderation group, we probe into the influence of AI. This involves exploring the perception of the credibility of these suggestions, the impact of knowing that the suggestion is generated by AI, and expectations for AI in future similar tasks. Each interview lasts approximately 30 minutes. With participants’ consent, we record the interviews and complete the transcription using automated tools and manual review by the first author. To avoid privacy leaks, the data associated with these interviews would not be publicly disclosed. For qualitative data, thematic analysis is applied to interview transcripts, following a structured coding process similar to our empirical analysis in Section 3.

A total of 50 participants took part in the first phase (25 in the human moderation group and 25 in the human-AI moderation group) and each completed 70 judgment tasks. Subsequently, 15 participants participated in semi-structured interviews (7 in the human moderation group and 8 in the human-AI moderation group). Through two phases, we obtain the quantitative results of participants under different experimental conditions, and gather their in-depth thoughts of the moderation process.

\section{Results}

This section presents the experimental results. Initially, we show the overall results of human and human-AI moderation groups in the judgment task to verify the balance of data distribution. Then, for RQ1, we investigate the judgment variations of the human moderation group under different perturbations. For RQ2, we analyze the judgment changes of the human-AI moderation group under different conditions. Finally, we compare the differences between the two modes to uncover the potential influence of AI assistance. 

Table \ref{tab:distribution} shows the number of tasks completed by each group. The distributions of the two groups are similar: the initial version accounts for approximately 13\%. Most perturbed versions are concentrated within the range of 8\% to 10\%, while the proportion of the statement order is lower (around 5\%). The reason is that the perturbation of statement order just generates one version by adjusting the order of two parties' statements, while other perturbation types involve operations on both the consumer and merchant sides. Besides, we perform a power analysis on the experimental data using G$\ast$Power \cite{Faul_119}. Because our comparisons are based on paired results derived from the same sample ID under different conditions, we use a paired \textit{t}-test method. Assuming the effect size Cohen's $d_z=0.50$, significance level $\alpha = 0.05$, and a statistical power $(1 - \beta) = 0.95$, the required size of paired samples is 54. Paired samples completed in both groups meet the threshold, ensuring sufficient statistical power for subsequent analysis. To avoid subjective biases of participants, we fit a mixed-effects model to the judgments in the human moderation group under the initial condition \cite{Blanchet_120}. User ID and initial sample ID are set as random intercepts, and the choices ``\textit{Extremely/support the merchant}'', ``\textit{Neutral}'', and ``\textit{Extremely/support the consumer}'' are encoded as -1, 0, and 1, respectively. The model intercept is not significant ($\beta = -0.159$, $SE = 0.082$, $p = 0.052$), indicating that participants do not exhibit a significant bias towards either merchants or consumers. For random effects, the variance at the user level is small ($SD = 0.31$), suggesting that participants did not exhibit notable individual differences; the sample-level variance ($SD = 0.62$) shows that different initial samples introduced some fluctuation in moderation, but it does not reach statistical significance.

\begin{table}[t]
\centering
\small
\setlength{\tabcolsep}{1pt}
\caption{Distribution of completed samples between two groups under different perturbation conditions.}
\label{tab:distribution}
\begin{tabular}{cccc}
 \hline
\parbox[c]{1.8cm}{\centering \textbf{Power Manifestation}} & \parbox[c]{1.8cm}{\centering \textbf{Human moderation group}} & \parbox[c]{1.8cm}{\centering \textbf{Human–AI moderation group}}  & \parbox[c]{1.7cm}{\textbf{$\Delta$ Percentage}} \\ \hline
Initial version    & 224 (12.80\%) & 227 (12.97\%) & +0.17\% \\
Legitimate claim   & 160 (9.14\%)  & 156 (8.91\%)  & -0.23\% \\
Authority citation & 161 (9.20\%)  & 159 (9.09\%)  & -0.11\% \\
Punishment threat  & 165 (9.43\%)  & 162 (9.26\%)  & -0.17\% \\
Compensation       & 159 (9.09\%)  & 163 (9.31\%)  & +0.22\% \\
Expert knowledge   & 156 (8.91\%)  & 158 (9.03\%)  & +0.12\% \\
Group preference   & 156 (8.91\%)  & 164 (9.37\%)  & +0.46\% \\
Statement order    &  86 (4.91\%)  &  84 (4.80\%)  & -0.11\% \\
Expression tone    & 162 (9.26\%)  & 164 (9.37\%)  & +0.11\% \\
Choice trap        & 164 (9.37\%)  & 157 (8.97\%)  & -0.40\% \\
Length difference  & 157 (8.97\%)  & 156 (8.91\%)  & -0.06\% \\
 \hline
\end{tabular}
\end{table}

\subsection{RQ1: Human moderation shows a tendency to support the powerful party under most perturbations}
In human moderation, we find that five power manifestations can significantly affect moderators' judgments (see the Appendix Table \ref{tab:within_groups}), including legitimate claim, punishment threat, compensation, expert knowledge, and length difference, triggering moderators' biases towards supporting the powerful party (the merchant). Moreover, although other perturbations do not reach statistical significance, distribution changes and interview feedback suggest that they may also impact moderators' judgments.

\begin{figure}[ht]
  \small
  \centering
  \includegraphics[width=1\linewidth]{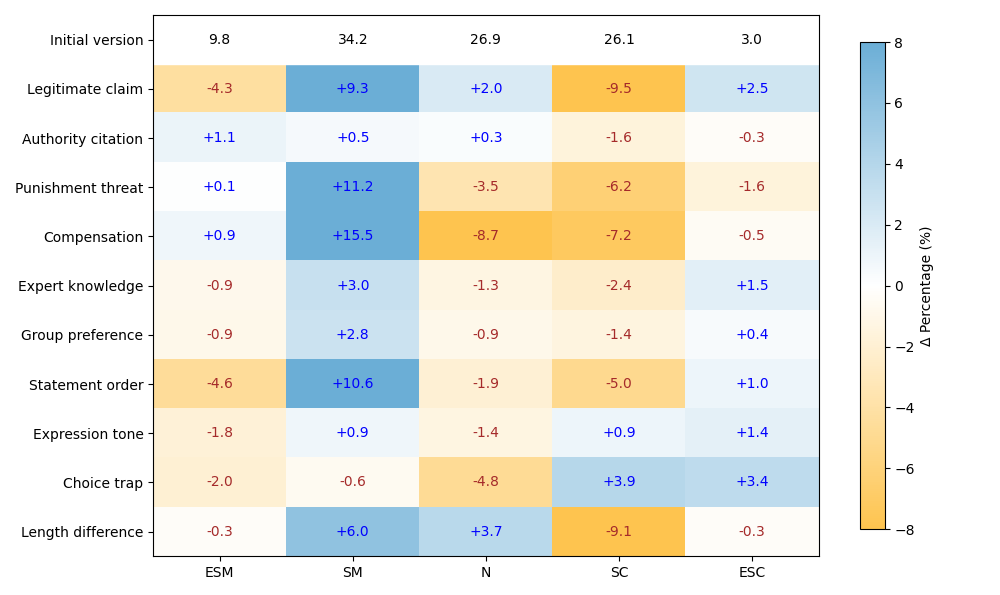}
  \caption{Choice variations without AI (initial version vs. perturbed version). For presentation convenience, ``Extremely support the merchant'', ``Support the merchant'', ``Neutral'', ``Support the consumer'', and ``Extremely support the consumer'' are abbreviated as ``ESM'', ``SM'', ``N'', ``SC'', and ``ESC'' respectively.}
  \Description{The figure is a heatmap showing the choice variations without AI (initial version vs. perturbed version) across five choices. For presentation convenience, ``Extremely support the merchant'', ``Support the merchant'', ``Neutral'', ``Support the consumer'', and ``Extremely support the consumer'' are abbreviated as ``ESM'', ``SM'', ``N'', ``SC'', and ``ESC'', respectively. Each row represents a version (e.g., legitimate claim, authority citation, punishment threat), and each cell shows the percentage point increase (blue) or decrease (orange) compared to the initial version, with the intensity of shading indicating the magnitude of change. }
\label{fig:human_moderation}
\end{figure}

\begin{figure}[ht]
  \small
  \centering
  \includegraphics[width=1\linewidth]{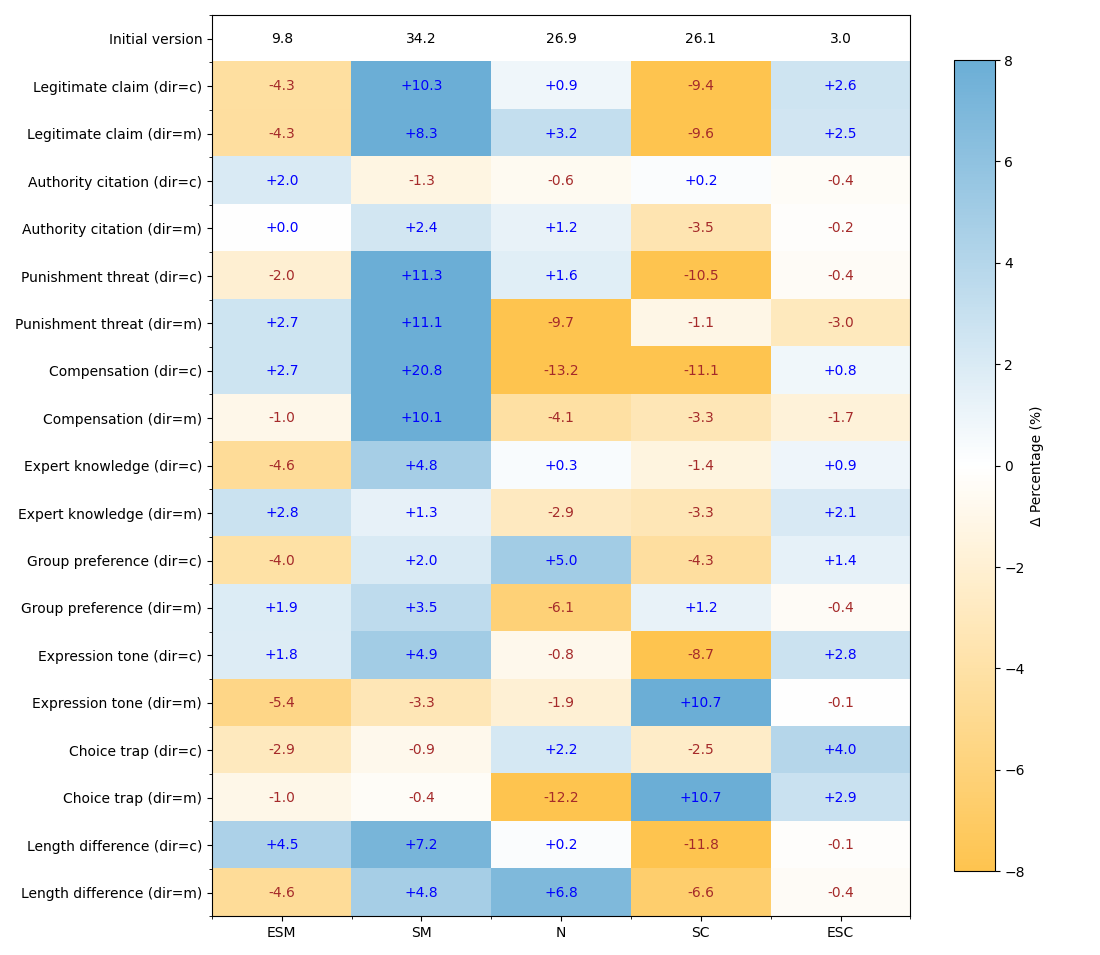}
  \caption{Choice variations without AI (initial version vs. perturbed version under two parties, using the same representations as Figure 3).}
  \Description{The figure is a heatmap showing the choice variations without AI under two parties (initial version vs. perturbed version, using the same representations as Figure 3). For presentation convenience, ``dir=c'' and ``dir=m'' denote perturbation on the consumer party and perturbation on the merchant party, respectively.}
\label{fig:human_moderation_side}
\end{figure}

\textit{Legitimate claim bias}. Human moderation presents a significant legitimate claim bias ($t(92)=2.284$, $p=0.025$, $d_z=0.237$), where moderators tend to support the powerful party (the merchant). As shown in Figure \ref{fig:human_moderation}, the proportions of choosing ``\textit{Support the merchant}'' and ``\textit{Neutral}'' increase by 9.3\% and 2.0\%, respectively, while that of ``\textit{Support the consumer}'' decreases by 9.5\%. This result indicates that when platform rules are present in the statement, moderators exhibit a significant bias towards supporting the merchant. Furthermore, the effect remains regardless of which party's statement contains the legitimate claim perturbation (Figure \ref{fig:human_moderation_side}), suggesting the legitimate claim itself serves as an independent factor. As reflected in the interview analysis, laws or rules are treated as credible signals. Some participants mentioned that laws or platforms' rules are ``\textit{formulated based on practical applications and should be followed}'' and ``\textit{reasonable and verifiable.}'' These responses suggest that the persuasiveness of the legitimate claim not only stems from the embedded information but also from the sense of trust brought by legitimacy.

\textit{Punishment threat bias}. Human moderation exhibits a significant punishment threat bias ($t(89)=4.375$, $p<0.001$, $d_z=0.461$), where moderators favor supporting the powerful party (the merchant). The proportion of choosing ``\textit{Support the merchant}'' increases by 11.2\%, while those of ``\textit{Support the consumer}'' and ``\textit{Extremely support the consumer}'' decrease by 6.2\% and 1.6\%, respectively. The trend remains regardless of which party's statement contains the punishment threat perturbation. According to the interview analysis, moderators believe that merchants are more capable of enforcing punishments and thus view threats from merchants as more credible. P3 said, ``\textit{The consumer's threat lacks the actual enforceability and is more like emotional expressions, while the merchant's threat is normal or official}.'' Therefore, moderators' interpretations of punishment threat statements are influenced by underlying social perceptions about power structures and enforcement capabilities of different parties. This makes threats from the powerful party more likely to be recognized as credible signals.

\textit{Compensation bias}. Human moderation shows a significant compensation bias ($t(97)=3.759$, $p<0.001$, $d_z=0.380$), where moderators tend to support the powerful party (the merchant). The proportion of choosing ``\textit{Support the merchant}'' increases by 15.5\%, while those of ``\textit{Neutral}'' and ``\textit{Support the consumer}'' decrease by 8.7\% and 7.2\%, respectively. Similarly, the phenomenon also remains regardless of which party's statement contains the compensation perturbation. The interview indicates the reasons for this phenomenon. Participants stated, ``\textit{The merchant's compensation demonstrated his/her sincerity and a sense of responsibility}'' (P2), while ``\textit{The discount request from the consumer is an unreasonable demand unrelated to the conflict fact}.'' (P7) When coming from the powerful party, it represents a positive attitude to resolve the conflict, while when coming from the weaker party, it is more prone to being questioned as opportunism.

\textit{Expert knowledge bias}. Human moderation presents a significant expert knowledge bias ($t(93)=2.357$, $p=0.021$, $d_z=0.243$), where moderators tend to support the powerful party (the merchant). Further analysis indicates a difference in terms of different parties. When the consumer's review inserts the expert knowledge, the proportion of choosing ``\textit{Extremely support the merchant}'' decreases by 4.6\%, while that of ``\textit{Support the merchant}'' increases by 4.8\%. When added in the merchant's response, the proportions of ``\textit{Extremely support the merchant}'' and ``\textit{Support the merchant}'' increased by 2.8\% and 1.3\%, respectively. This phenomenon suggests that moderators tend to trust professional expressions of merchants. Interviews also explain that the persuasiveness of expert knowledge is influenced by the party's identity or position. The professional terms used by merchants are often regarded as ``\textit{detail supplementation}'' or ``\textit{credible explanations}'', whereas similar expressions from consumers are easier to be interpreted as ``\textit{intentional ornamentation}'' and ``\textit{inconsistent with their level of knowledge.}'' P5 remarked directly, ``\textit{The professional terms cited by merchants are more persuasive, while consumers merely emphasize them deliberately to seek support}.'' Professional explanations from the powerful party tend to be regarded as credible, while those from the weaker party may be seen as imitated or motivated by improper intentions.

\textit{Length difference bias}. Human moderation shows a significant length difference bias ($t(93)=2.601$, $p=0.011$, $d_z=0.268$), where moderators tend to support the powerful party (the merchant). The proportion of choosing ``\textit{Support the merchant}'' increases by 6.0\%, while that of ``\textit{Support the consumer}'' decreases by 9.1\%. Furthermore, when consumers supply more content in their statements, the proportions of choosing ``\textit{Extremely support the merchant}'' and ``\textit{Support the merchant}'' increase by 4.5\% and 7.2\%, respectively. This suggests that moderators tend to support merchants even when consumers contain more content. The analysis also explains the reason behind it. Participants do not moderate based on the length, but make judgments by considering factors like content relevance and expression tone. Participants noted that longer content from merchants is perceived as ``\textit{detailed process}'' or ``\textit{more explanatory}'', whereas longer content from consumers contains ``\textit{irrelevant}'', ``\textit{emotional}'', or ``\textit{exaggerated}'' content. Therefore, the impact of length difference is also influenced by content relevance and sources.

Prior work has shown that differences in demographics can lead to moderators' variations. So we conduct a more in-depth analysis by considering moderators' two basic demographic characteristics—age and gender. The samples are divided into younger and older groups based on a median age split (26 years old), and into male and female groups based on gender, respectively. Paired \textit{t}-tests are then conducted between different groups, i.e., younger group VS older group, and male group VS female group. The results show that age and gender have different effects across different perturbations (see the Appendix). For age, both groups show significant effects under punishment threat and length difference perturbations. The younger group also exhibits significance under legitimate claim and expert knowledge, while the older group shows significance under authority citation and compensation perturbations. 
For gender, both groups are affected by the punishment threat perturbation, with a stronger effect size in the female group. The female group also has significance under legitimate claim and punishment threat perturbations, while the male group presents significance under authority citation, compensation, expert knowledge, and length difference perturbations. These results suggest that different groups vary in their sensitivity to different perturbations. Younger users tend to focus on the logical consistency of content, and older users are more sensitive to authority or interest-related information. Females are more likely to pay attention to risk or social norms, while males focus more on authority or expertise cues.

\textbf{Overall, human moderation shows five power-related biases towards supporting the powerful party.}

\subsection{RQ2: Power-related biases persist in human-AI moderation}
\subsubsection{Human-AI moderation shows a tendency to support the powerful party under most perturbations}
In human-AI moderation, we find that five power manifestations can affect moderators' judgments (see the Appendix Table \ref{tab:within_groups}), including legitimate claim, authority citation, punishment threat, compensation, and length difference, triggering biases towards supporting the powerful party (the merchant). Since the interview analysis results under different perturbations are similar to those of the human moderation group, the following will not be presented to avoid repetition. Although other perturbations do not reach statistical significance, the proportion of choosing ``\textit{Neutral}'' shows a declining trend, suggesting that these manifestations promote moderators to shift judgments from neutrality to clear stances. 

\begin{figure}[ht]
  \small
  \centering
  \includegraphics[width=1\linewidth]{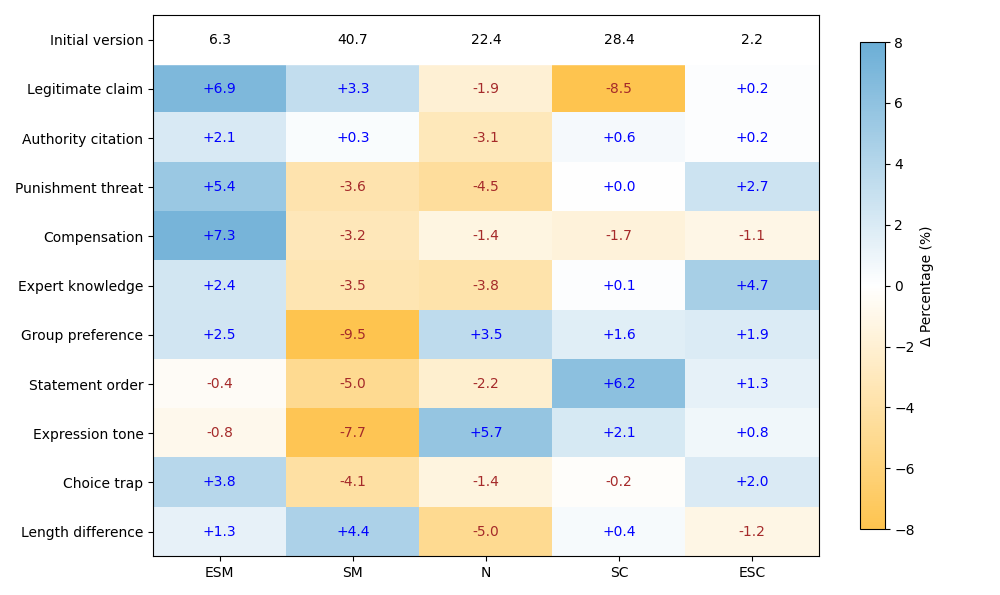}
  \caption{Choice variations with AI (initial version vs. perturbed version, using the same representations as Figure 3).}
  \Description{The figure is a heatmap showing the choice variations with AI (initial version vs. perturbed version, using the same representations as Figure 3).}
\label{fig:human_AI_moderation}
\end{figure}

\begin{figure}[ht]
  \small
  \centering
  \includegraphics[width=1\linewidth]{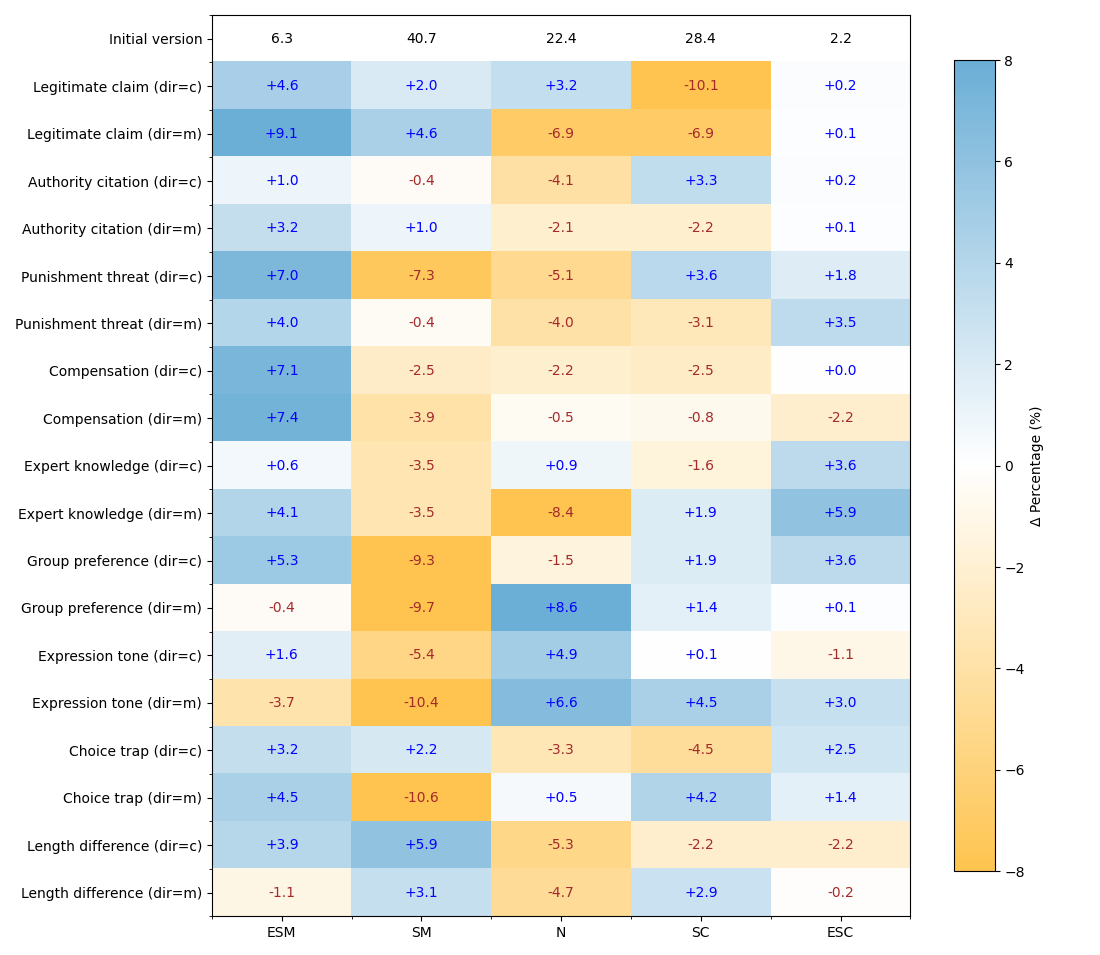}
  \caption{Choice variations with AI under two parties(initial version vs. perturbed version, using the same representations as Figure 4).}
    \Description{The figure is a heatmap showing the choice variations with AI under two parties (initial version vs. perturbed version, using the same representations as Figure 4).}
\label{fig:human_AI_moderation_side}
\end{figure}

\textit{Legitimate claim bias}. Similar to human moderation, human-AI moderation presents a significant legitimate claim bias ($t(94)=4.375$, $p<0.001$, $d_z=0.521$), where moderators tend to support the powerful party (the merchant). As shown in Figure \ref{fig:human_AI_moderation}, the proportions of choosing ``\textit{Extremely support the merchant}'' and ``\textit{Support the merchant}'' increase by 6.9\% and 3.4\%, respectively, while those of ``\textit{Neutral}'' and ``\textit{Support the consumer}'' decrease by 1.9\% and 8.5\%, respectively. Furthermore, the effect remains regardless of which party's statement contains the legitimate claim perturbation (Figure \ref{fig:human_AI_moderation_side}). 

\textit{Authority citation bias}. Compared to human moderation, human-AI moderation presents a new bias - authority citation bias, with a significant level ($t(96)=4.375$, $p=0.003$, $d_z=0.309$), which indicates moderators tend to support the powerful party (the merchant). The proportion of choosing ``\textit{Support the merchant}'' increases by 3.0\%, while that of ``\textit{Neutral}'' decreases by 3.1\%. Furthermore, when the perturbation is inserted into the consumer's review, the proportion of ``\textit{Support the consumer}'' increases by 3.3\%, while that of ``\textit{Neutral}'' decreases by 4.1\%. Conversely, when similar content is added to the merchant's response, the proportions of ``\textit{Extremely support the merchant}'' and ``\textit{Support the merchant}'' increased by 3.2\% and 1.0\%, respectively (Figure \ref{fig:human_AI_moderation_side}). This suggests that moderators tend to trust the party that contains authority citations. The interview analysis further indicates that the authority citation is regarded as a reliable cue. Even if the cited website is not verified, its existence can enhance moderators’ perceived reliability. P8 noted, ``\textit{When I see citations, given the verification by AI, I perceive them as more trustworthy. Even if the link cannot be accessed, it may be visible on a particular platform.''} Therefore, moderators' judgments rely on only logical expressions but are also heuristically influenced by authoritative citations.

\textit{Punishment threat bias}. Human-AI moderation exhibits a significant punishment threat bias ($t(96)=2.015$, $p=0.047$, $d_z=0.205$), where moderators tend to support the powerful party (the merchant). The proportions of choosing ``\textit{Extremely support the merchant}'' and ``\textit{Extremely support the consumer}'' increase by 5.4\% and 2.7\%, respectively, while those of ``\textit{Support the merchant}'' and ``\textit{Neutral}'' decrease by 3.6\% and 4.5\%, respectively. When the perturbation is added to the consumer's reviews, moderators tend to support the merchant. 

\textit{Compensation bias.} Human-AI moderation shows a significant compensation bias ($t(99)=3.718$, $p=0.001$, $d_z=0.358$), where moderators tend to support the powerful party (the merchant). The proportion of choosing ``\textit{Extremely support the merchant}'' increases by 7.3\%, while other options show a declining trend. Similarly, the effect remains consistent regardless of which party's statement contains the compensation perturbation. 

\textit{Length difference bias}. Human-AI moderation presents a significant length difference bias ($t(98)=2.473$, $p=0.012$, $d_z=0.257$), where moderators tend to support the powerful party (the merchant). The proportions of choosing ``\textit{Extremely support the merchant}'' and ``\textit{Support the merchant}'' increase by 1.3\% and 4.4\%, respectively, while that of ``\textit{Neutral}'' decreases by 5.0\%. When there is more content in the consumer's reviews, moderators tend to support the merchant. When merchants have more responses, moderators are prone to support consumers.

\begin{figure}[ht]
  \small
  \centering
  \includegraphics[width=1\linewidth]{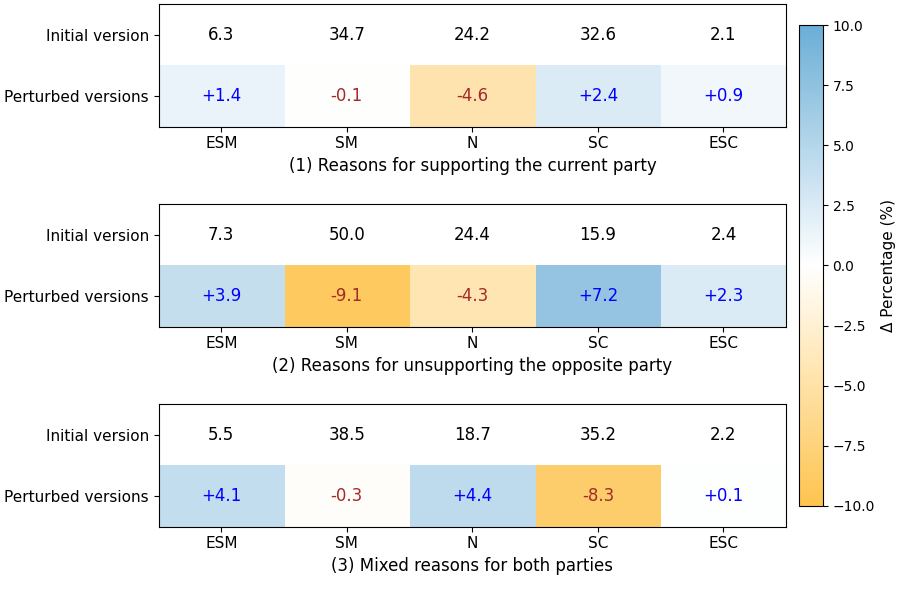}
  \caption{Choice variations across three perspectives of reasons in AI-generated suggestions (initial version vs. perturbed versions, using the same representations as Figure 3).}
  \Description{The figure contains three panels comparing the choice variations across three perspectives of reasons in AI-generated suggestions (initial version vs. perturbed versions, using the same representations as Figure 3). Each panel represents a different reason perspective, including supporting the current party, unsupporting the opposite party, and mixed reasons for both parties. It shows the percentage change of all perturbed versions compared to the initial version.}
\label{fig:AI_suggestions}
\end{figure}

The perspective of what reasons are presented in AI’s suggestions can influence moderators' judgments. Specifically, the reasons for unsupporting the opposite party have a significant effect ($t(93)=-2.736$, $p=0.008$, $d_z=-0.514$). The proportions of ``\textit{Support the consumer}'' and ``\textit{Extremely support the consumer}'' increase by 7.2\% and 2.3\%, respectively, while those of ``\textit{Support the merchant}'' and ``\textit{Neutral}'' decrease by 9.1\% and 4.3\%, respectively (Figure \ref{fig:AI_suggestions}). This suggests that the reasons of this perspective can significantly shift moderators' support to the consumer. Conversely, the reasons for supporting the current party do not show a significant difference, with the proportion of ``\textit{Support the consumer}'' increasing by 2.4\% and that of ``\textit{Neutral}'' decreasing by 4.6\%. Similarly, mixed reasons for both parties do not reach significance. In this case, the proportions of ``\textit{Extremely support the merchant}'' and ``\textit{Neutral}'' increase by 4.1\% and 4.4\%, respectively, while that of ``\textit{Support the consumer}'' decreases by 8.3\%. Interviews explain this. Most participants reported that the reasons of unsupporting the opposite party can ``\textit{expose the logical issue directly}'' and encourage them to ``\textit{think from the opposite perspective}''. However, participants said the reasons for supporting the current party are ``\textit{similar to my thoughts,}'' while mixed reasons are regarded as ``\textit{too much information and lacking focus.}''

\textbf{Therefore, human-AI moderation also exhibits power-related biases towards supporting the powerful party. Besides, the perspective for unsupporting the opposite party in AI-generated suggestions can motivate moderators to support the weaker party.}

\subsubsection{AI alleviates most power-related biases while amplifying a few} 
The experimental results demonstrate that the introduction of AI has different impacts on moderators (see the Appendix Table \ref{tab:between_groups}) under different perturbations. For most power-related biases existing in human moderation, the introduction of AI-generated suggestions alleviates moderators' biases in supporting the powerful party. However, under the legitimate claim condition, AI amplifies the existing bias.

\begin{figure}[ht]
  \small
  \centering
  \includegraphics[width=1\linewidth]{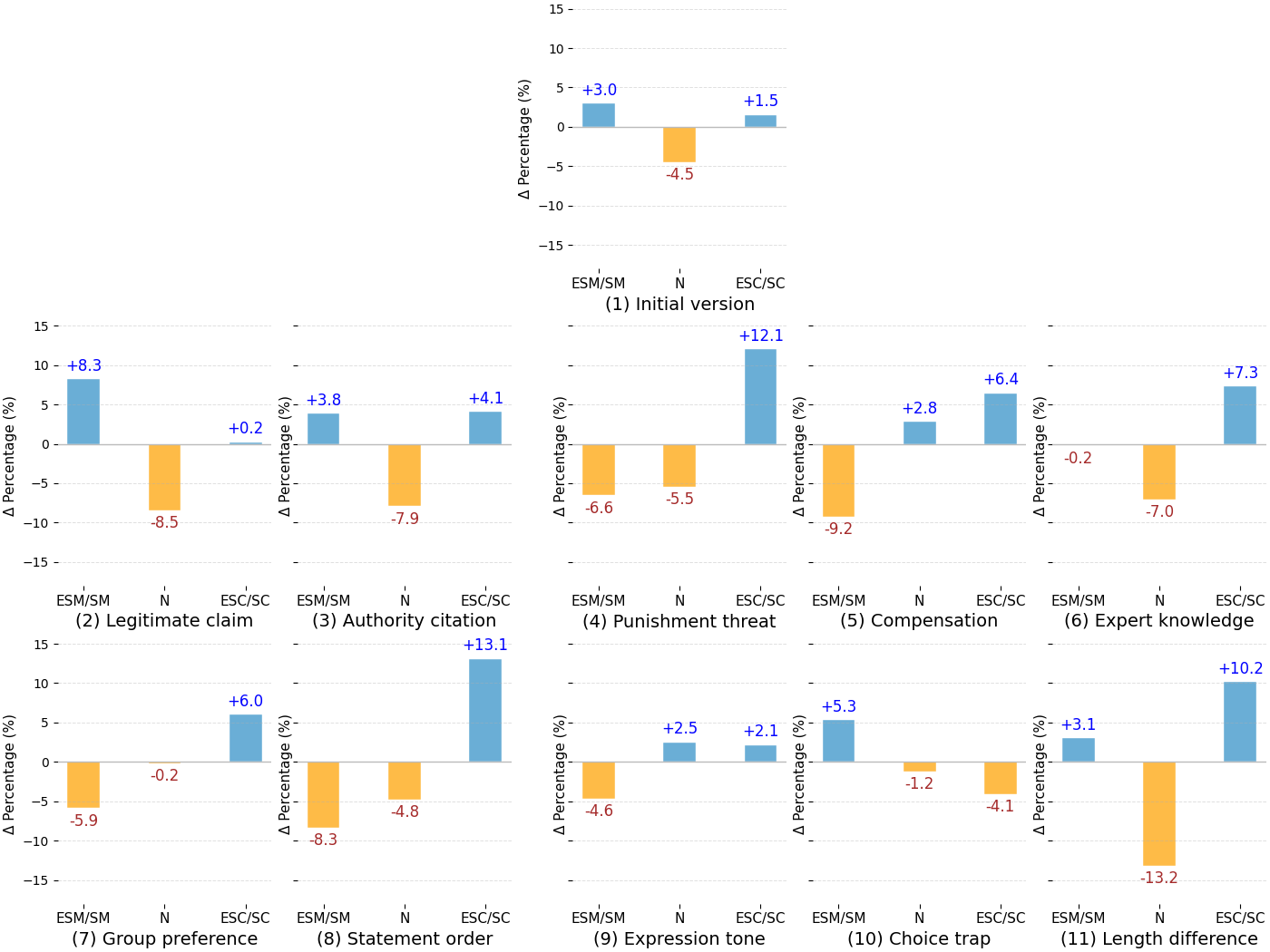}
  \caption{Choice variations with AI compared to without AI under different perturbed versions (using the same representations as Figure 3).}
  \Description{This figure contains eleven bar charts showing the choice variations with AI compared to without AI under different perturbed versions. Each subplot corresponds to one version (e.g., legitimate claim, authority citation, punishment threat) and shows the percentage point increase (blue) or decrease (orange) relative to human moderation. For presentation convenience, ``Extremely support the merchant'', ``Support the merchant'', ``Neutral'', ``Support the consumer'', and ``Extremely support the consumer'' are abbreviated as ``ESM'', ``SM'', ``N'', ``SC'', and ``ESC'' respectively. Positive and negative values are encoded both through bar direction and text labels to ensure accessibility without reliance on color.}
\label{fig:compare_two_modes}
\end{figure}

\textit{Alleviated biases.} The introduction of AI-generated suggestions alleviates several power-related biases in supporting the powerful party (the merchant), including punishment threat bias, compensation bias, expert knowledge bias, and length difference bias (Figure \ref{fig:compare_two_modes}). For the punishment threat perturbation, AI-generated suggestions show statistical significance ($t(86)=2.520$, $p=0.014$, $d_z=0.270$). The proportion of ``\textit{Extremely/Support the consumer}'' increases by 12.1\%, while that of ``\textit{Neutral}'' and ``\textit{Extremely/Support the merchant}'' decreases by 5.5\% and 6.6\%, respectively. This suggests that the introduction of AI effectively alleviates the effect of merchants and shifts moderators' support towards consumers. The interviews further explain this. P12 noted, ``\textit{The reasons point to threatening expressions and remind me that the judgment should be unbiased rather than emotional influence}.'' For the compensation perturbation, compared to human moderation, the proportion of ``\textit{Extremely/Support the consumer}'' in human-AI moderation increases by 6.4\%, while that of ``\textit{Extremely/Support the merchant}'' decreases by 9.2\%. Similar results are observed in expert knowledge and length difference perturbations. The interview analysis explains that AI-generated suggestions tend to be treated as reminders. When the expert knowledge from one party makes moderators confused, moderators can shift their focus to the conflict points identified by the AI, reducing the reliance on expert knowledge and identity. P10 said, ``\textit{The conflict points listed in LLM-generated suggestions enable me to directly compare the logic flaws in both parties' statements, rather than being swayed by professional terms.}''

\textit{Introduced and Amplified biases.} As mentioned in the previous section, human-AI moderation introduces a new authority citation bias. Besides, the introduction of AI-generated suggestions amplifies the legitimate claim bias in supporting the powerful party (the merchant). Compared with the results of human moderation, the proportion of ``\textit{Extremely/Support the merchant}'' in human-AI moderation increases by 8.3\%, while that of ``\textit{Neutral}'' drops by 8.5\%. This indicates that AI reduces neutral stances and strengthens the support for merchants, thus amplifying the tendency to favor the powerful party. The feedback analysis indicates that moderators believe that the rules cited by AI are credible. P9 supplemented, ``\textit{Since AI has cited rules, it indicates that they are reliable. So the claim by the merchant is more trustworthy.}''

\textit{Perceptions of AI roles}. The interview analysis indicates, participants generally regard AI-generated suggestions as a reference. When their judgments are ambiguous or neutral, AI could help them form explicit stances quickly, saving time for thinking and reading. As P10 stated, ``\textit{The reason list supplements my thoughts from different perspectives. When facing ambiguous or lengthy statements from both parties, AI can quickly identify core conflict points}.'' However, when AI-generated results are inconsistent with personal decisions, participants also express a cautious attitude. Some participants question AI's results and insist on their judgments. P12 noted, ``\textit{Sometimes AI also makes mistakes and lacks human empathy in ambiguous situations without standard answers}.'' 

\textit{Expectations of AI assistance}. Our analysis also identifies users' expectations for future improvements: participants hope that AI can offer more targeted, analytical functions to better support their moderation, including providing legal evidence, highlighting conflict points, and simulating conflict processes, etc. For instance, P8 pointed out, ``\textit{AI should directly present details of relevant laws or rules, reducing burden of manual checks.}'' P11 emphasized the need for a clearer comparative framework, ``\textit{AI should highlight logical contradictions in both parties' statements by changing colors or bolding texts.}'' P13 further suggested dynamic scenario support, ``\textit{AI can simulate conflict prothe cess between both parties' stances to help understand the scenario more intuitively.}''

\textbf{Overall, although AI assistance alleviates most biases of human moderation (four alleviated and one eliminated), it also amplifies a few (one introduced and another amplified).}

\section{Discussion}
Based on the above results, we identify several biases in power-asymmetric conflict moderation. In this section, we will discuss the potential mechanisms and effects of these biases. Based on that, multiple implications are proposed to inspire future research and design.

\subsection{Findings}
\subsubsection{Moderation biases stem from the perception of power asymmetry and heuristic thinking}
Our experimental results show that in both human and human-AI moderation, when conflict content contains perturbations including legitimate claim, punishment threat, compensation, and length difference, moderators tend to support the powerful party (the merchant). This indicates that bias in power-asymmetric conflicts is not incidental but rooted in long-standing social opinions. In other words, the bias does not arise from the labels ``consumer'' or ``merchant'', but from the potential perception that merchants hold power and consumers are weaker in social interactions. However, the power relation between merchants and consumers is not fixed but can shift with factors such as evidence granularity, platform rules, or public discourse \cite{Helberger_101}. For instance, when public opinion favors supporting consumers, moderators' perceptions that merchants are more powerful may be reduced. Such a perception difference can lead to the same language cues being interpreted differently on roles or contexts.

Power-related biases may arise from two main factors. First, default social perception of power asymmetry. The party with greater resources and interpretive authority is often viewed as more reliable. On e-commerce platforms, this perception privileges merchants, who are usually associated with symbols like brand reputation, service quality, and contract terms. When they explain or resolve conflicts with a service-oriented tone, their behaviors tend to be deemed legitimate. This reflects the System Justification Theory in social psychology, which suggests that individuals tend to hold existing social orders to reduce cognitive dissonance, even if it may harm personal interests \cite{Jost_88}. This perception is deeply ingrained and rarely questioned. Second, moderators' judgments tend to depend on heuristic thinking under constraints of limited information, heavy workload, and unverifiable evidence. It is typically a peripheral route processing according to the Elaboration Likelihood Model \cite{Petty_94}. In moderating consumer-merchant conflicts, moderators are volunteers and lack sufficient time to conduct a comprehensive examination of the evidence presented by both parties, so expressions like rules or punishment threats are perceived as surface cues by moderators for judgment. When default perceptions and heuristic thinking work together, the same language can be interpreted differently depending on the role: in merchants' statements, they are perceived as conveying the meaning of rules compliance, formality, and sincerity; while in consumers' statements, they are likely to be regarded as unreasonable, emotional, or manipulative. 

\subsubsection{The perception of AI-generated suggestions: a double-edged sword}
Our research also reveals a double-edged sword role of the moderator's perception of AI: it alleviates most biases that exist in human moderation, but also amplifies a few. Most power-related biases are mitigated to some extent under human-AI moderation, decreasing the supporting tendency towards the powerful party. The reasons might include two aspects. First, the suggestions help moderators quickly grasp core conflict points, reducing their reliance on heuristic cues like ``\textit{longer texts are more credible}''. Second, when moderators are neutral or lean towards merchants, AI-generated suggestions provide an alternative chance to reintroduce consumers' perspectives into consideration, helping them break away from the thinking of default social perceptions and pay more attention to consumers' statements. The situation is different in conflicts where the power relations are comparable, such as conflicts in online communities due to differences in contributions or conflicts in gaming communities caused by different understandings of rules \cite{Kittur_121}. There may be minor differences in action ability among parties, so moderators tend to focus more on the content itself rather than the party's identity or power. In these scenarios, the main role of AI lies in helping moderators improve the moderation efficiency \cite{Yu_122}.

The effect of perceptions of AI is not always positive. Our study indicates that human-AI moderation amplifies the legitimate claim bias towards the powerful party and introduces a new authority citation bias. This may stem from moderators' ``legitimizing'' effect on AI-generated content. When they encounter statements like rules in AI-generated suggestions, they potentially believe that the content is reliable. This effect creates a cumulative impact when moderators see the powerful party's statements containing the legitimate claim or authority citation, and is further strengthened through heuristic thinking, enhancing trust in the powerful party. Our interview feedback echoes this conclusion, ``\textit{When I encounter citations from specific websites, given the verification by AI, I perceive them as more trustworthy}.'' In essence, AI-generated suggestions are not only perceived as references but as a ``completed verification,'' weakening moderators' motivation to re-examine. Besides, the perspective of what reasons are presented in AI-generated suggestions also shapes judgments. When suggestions adopt the perspective of unsupporting the opposite party, moderators tend to support the weaker party. The possible reason is that such unsupporting reasons highlight logic flaws in the opposing party's statements, with a sense of accusation and negation, thus more likely to change moderators' judgments. Conversely, other presentations (perspectives for supporting the current party and mixed perspectives) containing positive arguments are less persuasive and insufficient to significantly change the original judgment.

These findings echo discussions in HCI regarding human-AI collaboration. People are often influenced by ``illusion of validity'' from LLMs \cite{Danry_92, Kaur_93}. In scenarios like autonomous driving, seemingly plausible AI reminders are often mistakenly regarded as reliable \cite{Rosbach_72}. In content understanding, deceptive explanations by LLMs can even enhance people's trust in fake news compared to authentic explanations \cite{Danry_92}. Our findings further extend current research. In power-asymmetric conflict moderation, AI does not merely serve the function of information assistance. Instead, its existence and presentation forms also shape moderators' bias patterns. This double-edged sword effect highlights that future design of AI-assisted systems should extend beyond efficiency and incorporate mechanisms that alleviate moderation biases, thereby mitigating the impact of moderators' ``legitimizing'' effect on AI.

\subsubsection{Commonalities and individual differences in power-asymmetric conflict moderation}
This study shows that moderators' judgment of power-asymmetric conflicts follows a certain common pattern. They generally begin with looking at the consumer’s queries and check whether the merchant has addressed each conflict point. Then formulate an initial judgment by integrating the richness of details in aspects like product quality, platform rules, and service demands. For example, when a consumer claims, ``\textit{The staff had a bad attitude, and the food tasted ordinary,}'' and the merchant responds with, ``\textit{All staff on duty that day have received training, and attached screenshots to prove the staff have been smiling throughout service. Besides, our food is prepared with standardized procedures...},'' then the consumer's claim ``\textit{bad attitude}'' appears insufficient due to insufficient detail. Conversely, the merchant provides targeted responses to conflict points of service attitude and food quality. Tone also impacts moderators' judgments. Objective and structured statements are deemed more persuasive, while emotional statements tend to trigger doubts. Overall, moderators tend to follow a specific pattern in moderation. They first check whether the responses match, then evaluate details and tone, and finally form a judgment.

However, moderators' judgments exhibit some individual differences, which are related to factors such as their experience, demographics, and values. First, identity and personal experience can affect moderation. Moderators not only play the role of third-party judges but are also consumers in daily lives. This two-fold identity can subtly influence judgments. P2 remarked,''\textit{I have personal experience of staying in hotels. They usually provide pick-up and drop-off services. It's unreasonable for this hotel to claim otherwise.}'' When P5 judged the same case, he stated, ``\textit{The primary function of a hotel is to provide accommodation. Pick-up and drop-off services are supplementary. If it is clearly communicated before booking that the service is not available, there's no issue.}'' This aligns with Instance-Based Learning Theory, which suggests that individuals adjust strategies based on the similarity between the current situation and past experience \cite{Gonzalez_old75}. Second, judgment can be influenced by human values. Moderators who prioritize the protection of the weaker party tend to support consumers, while those who emphasize order and institutional stability tend to support merchants. As P7 noted, ``\textit{These consumers' reviews should be posted because others can gain insights into the service details of the merchant},'' while P9 said, ``\textit{Reviews must be grounded in facts and should not be emotional complaints.}'' According to Schwartz's value model, values serve as guiding principles in individual behavior. When statements align with moderators' value system, they tend to view them as ``\textit{reasonable}'' or ``\textit{acceptable}''. When the value claims in statements are inconsistent with their values, they tend to question or deny those statements. Third, differences in moderators' imaginations on AI can influence moderation. These imaginations stem not from understandings of how AI actually works, but from the mental models of AI formed by moderators through daily technology use. Such mental models of AI refer to people's internal beliefs about what AI can do, how it makes judgments, and why it produces certain outputs \cite{Gero_100, Mohanty_123}. Prior work shows that AI literacy shapes these mental models of AI. A higher level of AI literacy enables appropriate reliance on automated systems, while lower literacy exhibits over-reliance or excessive suspicion \cite{Horowitz_old74, Ehsan_99}. This point is reflected in our interview. For example, P9 stated, ``\textit{As an important information assistant, LLM-driven AI now solves many problems in my daily work.}'' Conversely, P12 commented, ``\textit{AI just grabs content from the internet and sometimes even fabricates it. It’s not reliable.}'' Thus, varying levels of technical literacy influence how moderators imagine and trust AI, which in turn affects their moderation judgments.


 \subsection{Design implications}
Based on the above findings, we propose the following design implications.
\begin{itemize}
    \item \textbf{Alleviate moderation biases stemming from default power asymmetry}. In power-asymmetric conflicts, moderation should pursue not only efficiency but also alleviate moderation biases. Two future directions are suggested based on our research. First, anonymization and symmetrical information presentation can be considered to reduce biases associated with roles. Conflicts between consumers and merchants can be reframed as ``\textit{Subject A}'' and ``\textit{Subject B,}'' with statements presented in a standardized format such as ``\textit{Assertion/Evidence/Claim/Other Language Cues.}'' Automated sentence processing can further distill key information to minimize potential biases related to text length or rhetorical styles. This strategy echoes prior HCI research regarding the use of interface design to alleviate heuristic biases \cite{Hutson_77, Lee_78, Shu_125}. Second, bias reminders and counterfactual comparison mechanisms can be explored to strengthen moderators' meta-awareness. When language cues like compensation or punishment threat appear in the powerful party's statements, the system can generate two summaries with and without these cues to highlight the potential influence of judging from surface signals. The approach aligns with principles of explainability and transparency emphasized in the content moderation field \cite{Jhaver_76}. These features can help mitigate moderation biases stemming from default power asymmetry.
    \item \textbf{Control the ``legitimizing'' effect of AI-generated content on moderators}. Moderators often perceive AI-generated suggestions containing authority citation or rule details as verified evidence, reflecting a ``legitimizing'' effect. Future moderation systems are suggested to regulate this influence. First, uncertainty annotations like confidence intervals and multiple interpretations can be leveraged to replace deterministic expressions in systems, prompting moderators to think critically \cite{Miller_90, Lu_91}. Second, transparency can be enhanced through evidence chain tracing to prevent moderators' misinterpretation of AI outputs as ``\textit{completed verification.}'' For instance, add ``\textit{to be verified}'' indicators and visually depict evidence sources or reasoning paths. Similar designs have been proven effective in reducing over-reliance on AI in medical diagnosis \cite{Bussone_79}, aligning with the call for moderate reliance in HCI \cite{Lee_80}. These features do not undermine AI’s value but support balanced judgments that combine efficiency and alleviate moderation biases.
    \item \textbf{Improve moderation by incorporating both commonality and individuality}. Moderators under power-asymmetric conflicts show similar moderation patterns but also exhibit individual variances, indicating that systems should balance consistency with diversity. First, structured support can standardize core procedures. Our research shows moderators typically follow a sequence of ``matching responses to conflict points, verifying details, assessing tone.''  Introducing interactive checklists in moderation systems can guide this process by ensuring each conflict point is addressed. Second, systems should consider personal experiences and value orientations. Prior work also confirms that individual values shape moderation \cite{Jiang_83}. Therefore, functions like experience annotation (document past experiences) and configurable value-orientation settings (e.g., ``\textit{emphasis on norms}'') can capture these differences. Besides visualizations of distribution proportion, the group intelligence can also aggregate orientations using weighting algorithms, while measures like delayed or anonymous display can mitigate conformity effects \cite{Ueda_69, Dolin_81}. Third, it suggests considering variations in AI literacy. Hierarchical explanation interfaces can provide both concise and detailed justifications. Moreover, educational prompts upon initial use (e.g., ``\textit{the suggestions are only for reference!}'') and simulation practice on historical conflicts can further help moderators with lower AI literacy in shaping critical thinking.
\end{itemize}

\subsection{Generalizability}
This section discusses the generalizability of our work in terms of research findings and experimental design. Focusing on moderation bias in online power-asymmetric conflicts, we leveraged the consumer-merchant conflicts as a representative scenario, but the findings can be generalized into other contexts. First, the tendency of third-party moderators to support the powerful party is not limited to this scenario. In medical disputes, arbitrators might tend to accept doctors' interpretations due to their professional authority \cite{Bussone_79}. In education, evaluators may support teachers because of their knowledge and institutional standing \cite{Han_89}. These phenomena, similar to consumer-merchant conflicts, show that power asymmetry has a universal impact on third-party moderators. Thus, our findings can contribute understanding of bias patterns in these contexts. Second, the ``legitimizing'' effect of AI-generated content is also reflected in other decision-making scenarios. For example, doctors may rely on LLMs' outputs to reduce workload but overlook patient differences, and HR personnel amplify gender and racial biases when using LLMs for resume screening \cite{Dena_84}. These show that AI-generated content containing plausible language may exacerbate misjudgments, offering implications for moderation and broader human–AI collaboration. 

Moreover, our experimental design demonstrates generalizability and can be applied to other studies on human/AI bias and conflict judgments. First, the intervention approach used to examine bias is not limited to online consumer-merchant conflicts, but can be transferred to scenarios like workplace negotiations, community governance, and platform mediation. Second, the Wizard-of-Oz design (researchers acting as the ``AI'') eliminates the differences in models' outputs, allowing us to focus on moderators' bias patterns rather than technical constraints. This offers a typical framework for studying how people perceive AI-generated content and can inform the design of AI-assisted systems independent of specific models.

\section{Limitations and Future Work}
This work has the following limitations. First, our study leverages consumer-merchant conflicts as a representative scenario of power asymmetry. While this scenario offers accessible data and a standardized resolution process, just the one scenario may limit the generalizability and reliability of our findings. Other representative contexts like workplace negotiations, community governance, or medical disputes involve multiple rounds of interaction, implicit power dynamics, and moral considerations, which may trigger different biases. Future research should validate the findings in more contexts. Furthermore, our data was obtained from Dianping, a Chinese local life service platform, with all participants from China. While this ensures contextual consistency, it limits cross-cultural generalizability. Prior studies have shown that content moderation can be influenced by cultural differences, including identity characteristics, ideology, and language, etc. \cite{Andrew_46, Haimson_37, Nogara_48}. These findings imply that the moderation biases identified in our study could not be generalized to all cultures. So future studies should be carried out in diverse cultures to examine the universality and boundaries of the bias mechanisms uncovered in this work.

Besides, our study does not measure the participants’ technical literacy level, which limits the ability to assess how it affects their imaginations of AI and moderation judgments. Future work could consider different levels of technical literacy among moderators to further explore how these differences shape their trust in AI during moderation. Moreover, to avoid technical differences, we adopt a Wizard-of-Oz method wherein researchers act as ``AI'' to provide suggestions, inevitably departing from real usage scenarios of AI models. Particularly, different LLMs' outputs vary significantly in tone, accuracy, hallucination frequency, and even minor changes in prompts can lead to different responses, affecting users' trust \cite{Cohn_95, Wester_96, Wang_98}. Suggestions from real LLMs may alter the manifestation of moderators' biases due to their expression style and output stability. More authoritative expressions may amplify biases, while inconsistent outputs among different LLMs could undermine moderators’ trust. Future research could integrate real AI models for comparison and the corresponding influence on moderation.

\section{Conclusion}
In this paper, we investigated the power-related biases of human and human-AI moderation in power-asymmetric conflicts. Through a mixed design experiment involving 50 participants, we found that both moderation modes exhibit several power-related biases towards supporting the powerful party. Although AI-generated suggestion alleviates most biases of human moderation, it also amplifies a few. Moreover, presenting the reasons of unsupporting the opposite party in these suggestions also prompts moderators to support the weaker party to a certain extent. These findings provide several insights into the design of power-asymmetric conflict moderation mechanisms and AI-assisted moderation systems. In future work, we will further explore the differences in moderation biases across different cultural contexts and explore moderation designs as well as human-AI collaboration strategies to mitigate these biases.

\begin{acks}
This research was supported by Meituan and National Natural Science Foundation of China (NSFC) under the Grant No. 62372113. Peng Zhang is a faculty of College of Computer Science and Artificial Intelligence, Fudan University. Tun Lu is a faculty of College of Computer Science and Artificial Intelligence, Fudan University, Shanghai Key Laboratory of Data Science, and Silver-X MOE Philosophy \& Social Sciences Laboratory, Fudan Institute on Aging.
\end{acks}

\bibliographystyle{ACM-Reference-Format}
\bibliography{conflict_base}

\appendix
\section{Appendix}

\begin{table}[h]
\centering
\footnotesize
\setlength{\tabcolsep}{1pt}
\caption{Between-group paired \textit{t}-test results for the human and human-AI moderation groups. *** indicates $p<0.001$, ** indicates $p<0.01$, and * indicates $p<0.05$. Cohen’s $d_z$ reports the effect size on the dependent variables.}
\label{tab:within_groups}
\begin{tabular}{ccccccccc}
 \hline
\multirow{2}{*}{\textbf{Power Manifestation}} & \multicolumn{4}{c}{\parbox[c]{2cm}{\centering \textbf{Human moderation group}}} & \multicolumn{4}{c}{\parbox[c]{2cm}{\centering \textbf{Human-AI moderation group}}} \\ 
 & \textbf{sd} & \textbf{t} & \textbf{p} & \textbf{$d_z$}   & \textbf{sd}  & \textbf{t} & \textbf{p} & \textbf{$d_z$}\\ \hline
Legitimate claim & 1.181  & 2.284 & \textbf{0.025(*)}   & 0.237 & 0.884
& 5.077 & \textbf{0 (***)}     & 0.521 \\
Authority citation &1.346  & 1.897 & 0.061       & 0.194 &  0.972
& 3.08 & \textbf{0.003(**)}  & 0.309 \\
Punishment threat & 1.12  & 4.375 & \textbf{0(***)}     & 0.461 & 1.058 & 2.015& \textbf{0.047(*)}   & 0.205\\
Compensation    &  1.303   & 3.759 & \textbf{0(***)}     & 0.380 & 1.008 & 3.718& \textbf{0.001(**)}  & 0.358\\
Expert knowledge &1.094    & 2.357 & \textbf{0.021(*)}   & 0.243 & 1.024 & 1.233& 0.22& 0.125\\
Group preference & 1.261    & 1.893 & 0.061       & 0.192 & 0.974
& 1.032 & 0.305& 0.104\\
Statement order & 1.284    & 1.966 & 0.053       & 0.225 & 1.144
& 0.486& 0.629& 0.054\\
Expression tone  & 1.216   & 1.215 & 0.228       & 0.130  & 1.021
& 1.187& 0.238& 0.119\\
Choice trap    &  1.369    & 0.377 & 0.707       & 0.039 &  1.03
& 1.577& 0.118& 0.16\\
Length difference & 1.289  & 2.601 & \textbf{0.011(*)}   & 0.268 &  1.016& 2.473 & \textbf{0.012(*)}   & 0.248\\
 \hline
\end{tabular}
\end{table}

\begin{table}[h]
\centering
\footnotesize
\setlength{\tabcolsep}{1pt}  
\caption{Paired \textit{t}-test results between the initial and perturbed versions across age groups (using the same representations as Table 4).}
\label{tab:age}
\begin{tabular}{ccccccccc}
 \hline
\multirow{2}{*}{\textbf{Power Manifestation}} &  \multicolumn{4}{c}{\textbf{Younger group}}  &\multicolumn{4}{c}{\textbf{Older group}} \\ 
 &  \textbf{sd} &\textbf{t} & \textbf{p} & \textbf{$d_z$}  &  \textbf{sd} &\textbf{t} & \textbf{p} & \textbf{$d_z$}\\ \hline
Legitimate claim   &  1.181
&3.224 & \textbf{0.002(**)} & 0.423 &  1.284
&1.382 & 0.176 & 0.234 \\
Authority citation &  1.443
&0.809 & 0.422      & 0.111 &  1.366
&2.088 & \textbf{0.043(*)} & 0.322 \\
Punishment threat  &  1.153
&4.045 & \textbf{0(***)}   & 0.572 &  1.136
&2.532 & \textbf{0.016(*)} & 0.416 \\
Compensation       &  1.303
&1.982 & 0.052      & 0.263 &  1.284
&3.169 & \textbf{0.003(**)} & 0.478 \\
Expert knowledge   &  1.421
&2.226 & \textbf{0.030 (*)}  & 0.297 &  1.225
&1.847 & 0.072 & 0.288 \\
Group preference   &  1.363
&0.791 & 0.432      & 0.107 &  1.253
&1.598 & 0.120 & 0.278 \\
Statement order    &  1.242
&1.675 & 0.102      & 0.268 &  1.306
&1.756 & 0.096 & 0.403 \\
Expression tone    &  1.421
&1.027 & 0.310      & 0.150 &  1.093
&1.391 & 0.172 & 0.223 \\
Choice trap        &  1.38
&1.558 & 0.126      & 0.216 &  1.428
&0     & 1.000 & 0     \\
Length difference  &  1.243&2.729 & \textbf{0.008(**)} & 0.349 &  1.238&2.640 & \textbf{0.013(*)} & 0.467 \\
 \hline
\end{tabular}
\end{table}

\begin{table}[h]
\centering
\footnotesize
\setlength{\tabcolsep}{1pt}
\caption{Paired \textit{t}-test results between the initial and perturbed versions across gender groups (using the same representations as Table 4).}
\label{tab:gender}
\begin{tabular}{ccccccccc}
 \hline
\multirow{2}{*}{\textbf{Power Manifestation}} & \multicolumn{4}{c}{\textbf{Male group}}  &\multicolumn{4}{c}{\textbf{Female group}} \\ 
 &  \textbf{sd} &\textbf{t} & \textbf{p} & \textbf{$d_z$}  &  \textbf{sd} &\textbf{t} & \textbf{p} & \textbf{$d_z$}\\ \hline
Legitimate claim   &  1.27
&1.147 & 0.256       & 0.152 &  1.212
&3.236 & \textbf{0.003(**)} & 0.518 \\
Authority citation &  1.327
&2.234 & \textbf{0.030(*)}   & 0.301 &  1.533
&0.866 & 0.392      & 0.135 \\
Punishment threat  &  1.176
&2.418 & \textbf{0.020(*)}   & 0.353 &  1.209
&3.125 & \textbf{0.004(**)} & 0.514 \\
Compensation       &  1.406
&3.289 & \textbf{0.002(**)}  & 0.411 &  1.212
&1.071 & 0.291      & 0.174 \\
Expert knowledge   &  1.319
&2.785 & \textbf{0.007(**)}  & 0.372 &  0.901
&1.387 & 0.173      & 0.217 \\
Group preference   &  1.387
&0.550 & 0.585       & 0.076 &  1.223
&0.970 & 0.338      & 0.153 \\
Statement order    &  1.414
&1.414 & 0.166       & 0.236 &  1.265
&1.318 & 0.201      & 0.275 \\
Expression tone    &  1.136
&1.257 & 0.215       & 0.180 &  1.372
&0.928 & 0.362      & 0.182 \\
Choice trap        &  1.59
&0.424 & 0.673       & 0.057 &  1.271
&0.905 & 0.371      & 0.149 \\
Length difference  &  1.252&2.446 & \textbf{0.018(*)}   & 0.357 &  1.265&0.577 & 0.567      & 0.094 \\
 \hline
\end{tabular}
\end{table}

\begin{table}[h]
\centering
\footnotesize
\setlength{\tabcolsep}{2pt}
\caption{Paired \textit{t}-test results comparing the human and human-AI moderation groups (using the same representations as Table 4).}
\label{tab:between_groups}
\begin{tabular}{ccccc}
 \hline 
\textbf{Power Manifestation} & \textbf{sd} & \textbf{t} & \textbf{p} & \textbf{$d_z$} \\ \hline
Initial version  &  0.76    & -0.449 & 0.654       & -0.045 \\
Legitimate claim&  0.815     & -0.523 & 0.602       & -0.056 \\
Authority citation &0.979     &  0.106 & 0.916       &  0.011 \\
Punishment threat & 0.886     &  2.520 & \textbf{0.014 (*)}   &  0.270 \\
Compensation     &0.932      &  1.626 & 0.107       &  0.164 \\
Expert knowledge  & 0.839     &  0.663 & 0.509       &  0.069 \\
Group preference  &  1.022    &  1.165 & 0.247       &  0.119 \\
Statement order   &  1.096    &  1.026 & 0.309       &  0.128 \\
Expression tone & 1.006      &  0.998 & 0.321       &  0.108 \\
Choice trap    &  1.093       & -0.716 & 0.476       & -0.075 \\
Length difference &  0.896    &  1.022 & 0.309       &  0.106 \\
 \hline
\end{tabular}
\end{table}
\end{document}